\documentclass[aps,prb,epsf,floatfix]{revtex4}
\usepackage{color}
\usepackage{epsfig}
\usepackage{amssymb}
\usepackage{subfigure}

\newcommand{\be}{\begin{equation}}
\newcommand{\ee}{\end{equation}}
\newcommand{\bea}{\begin{eqnarray}}
\newcommand{\eea}{\end{eqnarray}}
\begin{document}

\title{Entanglement signatures of the quantum phase transition
induced by a magnetic impurity in a superconductor
}

\author{P. D. Sacramento$^{1,2}$, P. Nogueira$^2$, V. R. Vieira$^{1,2}$ and V. K.
Dugaev$^{1,2,3}$}
\affiliation{$^1$Departamento de F\'isica and 
$^2$Centro de F\'{\i}sica das Interac\c{c}\~oes Fundamentais,\\
Instituto Superior T\'ecnico, Universidade T\'ecnica de Lisboa (TULisbon),\\
Av. Rovisco Pais, 1049-001 Lisboa, Portugal\\
$^3$Department of Mathematics and Applied Physics, Rzesz\'ow University of Technology,
Al. Powsta\'nc\'ow Warszawy 6, 35-959 Rzesz\'ow, Poland
}

\date{\today}

\begin{abstract}
The insertion of a magnetic impurity in a superconductor induces a first order
quantum phase transition as the coupling to the electronic spin density
increases. As the transition is crossed, a discontinuity is exhibited by
various quantities, like the total spin density, the total gap function and the
gap function at the impurity location. The location of other quantum phase
transitions have been detected by singularities in entanglement measures of the
system. In this work we show that the single-site and two-site von Neumann
entropies, the mutual information and the Meyer-Wallach measure show
discontinuities at the quantum phase transition. The negativity is less
sensitive to the transition. We study in detail these quantities as a function
of spin coupling and distance from the impurity center.
\end{abstract}
\vspace{0.3cm}
\pacs{PACS numbers: }
\maketitle


\section{Introduction and model}

There has been in recent years a very rich interplay between quantum
information and computation and the study of quantum many-body
systems\cite{amico}.

The concept of entanglement of quantum states --- which was introduced
by Schr\"odinger\cite{schroedinger}, already in 1935, at the time of
the EPR paradox\cite{epr}, and latter explored by Bell\cite{bell} to
establish his famous inequalities --- has been a crucial factor in
the impressive developments in the area of quantum information and
computation.
This made possible the distinction between quantum and
classical physics on an operational basis, leading to the experimental 
confirmation\cite{clauser, aspect, tittel} of quantum mechanics.
  
The development of the formalism and techniques used in the area
of quantum information and computation\cite{nielsen.chuang} has
occurred in parallel with developments in other scientific areas,
with each area being a source of inspiration to each other.
The study of quantum many-body systems and their correlations
and phase transitions, both at zero and non-zero temperatures,
provides an unambiguous example of that mutual influence.

One example is the development of the Matrix Product States (MPS)\cite{MPS} and 
of the Projected Entangled-Pair States (PEPS)\cite{PEPS}, which provide an
efficient calculation of the low-lying energy states of quantum systems, and the
reformulation\cite{ostlund.mps,rommer.mps.prb} of the density matrix renormalization group 
(DMRG)\cite{white.dmrg}, allowing to understand and overcome some of its limitations, 
associated to the boundary conditions in the system.

This interrelation has also been explored in the reanalysis of several
non-trivial exactly solvable models, under the new methods and concepts, using
the known ground state or correlation functions to calculate the different
information measures and to better understand the underlying physics.
These models are typically either one dimensional
spin\cite{latorre-2004-4,gu-2005-71} or electron
chains\cite{gu, goteborg},
solvable by the Bethe ansatz or the Jordan-Wigner and the Bogoliubov-Valatin
transformations, or effectively infinite dimensional models, where each
particle interacts equally with all the others\cite{vidal-2006-73,vidal-2006}.
Intermediate dimensions or extensions of these models have also been studied
usually using numerical methods.

For a bipartite system, in particular for a two qubit system, the entanglement
measures are well established and essentially equivalent. However, this is not
the case for systems with a larger number of components\cite{vedral-entropy}. The issue is even
complicated by the restrictions on how to share entanglement by more than two
particles, the so-called monogamy of entanglement\cite{ent-monogamy}. Besides the von Neumann 
and related quantities\cite{nielsen.chuang}, other information measures like the
concurrence\cite{wooters}, the negativity\cite{vidal} or the Meyer-Wallach\cite{meyer} and the 
generalized global entanglement measures\cite{miranda} have been considered.

In most discussions on entanglement, the particles constituting the systems are
considered distinguishable. The Schmidt rank, central in this analysis, assumes
this, for example. However, this is not the case for the quantum statistics,
obeyed by the fermions and bosons, and new methods have to be used\cite{amico}. The effect 
of the quantum statistics has been analyzed for free electrons\cite{lunkes05-2}
and bosons\cite{heaney06}, and 
for electrons in a BCS superconductor\cite{oh04} or with the so called
eta-pairing\cite{vedral04-2}.

In this work we consider the quantum phase transition induced by a magnetic
impurity inserted in a conventional superconductor 
\cite{sakurai,satori,salkola,morr1,morr2,review,abrikosov60,ramazashvili97},
and study it using some of those entanglement measures as a signature of the
quantum phase transition.
Consider a classical spin immersed in a two-dimensional $s$-wave conventional
superconductor. We use a lattice description of the system.
In the center of the system ($i=l_c=(x_c,z_c)$) we place a classical spin parametrized like
\be
\frac{\vec{S}}{S} = \cos \varphi \vec{e}_x + \sin \varphi \vec{e}_z
\ee
where $S$ is the modulus of the spin. We assume that the spin lies in the plane of the superconducting film
($x-z$ plane). 
The Hamiltonian of the system is given by
\bea
H &=& - \sum_{<i,j>,\sigma} t_{i,j} c_{i\sigma}^{\dagger} c_{j\sigma} - e_F \sum_{i\sigma}
c_{i\sigma}^{\dagger} c_{i\sigma} 
+ \sum_i \left( \Delta_i c_{i\uparrow}^{\dagger} c_{i\downarrow}^{\dagger}
+ \Delta_i^* c_{i\downarrow} c_{i\uparrow} \right) \nonumber \\
&-& \sum_{\sigma,\sigma'} J_{l_c} [ \cos \varphi c_{l_c \sigma}^{\dagger}
\sigma_{\sigma,\sigma'}^x c_{l_c \sigma'} 
+ \sin \varphi c_{l_c \sigma}^{\dagger} \sigma_{\sigma,\sigma'}^z c_{l_c \sigma'} ],
\eea
where the first term describes the hopping of electrons between different
sites on the lattice, $e_F$ is the chemical potential, the third term
is the superconducting $s$-pairing with the site-dependent order
parameter $\Delta _i$, and the last term is the exchange interaction of an electron at
site $i=l_c$ with the magnetic impurity.
The hopping matrix is given by
$t_{i,j}=t \delta_{j,i+\delta}$ where $\delta$ is a vector
to a nearest-neighbor site.
Note that both the indices $l$ and $i,j$ specify sites on a two-dimensional
system.
The indices $i,j=1,...,N$, where $N$ is the number of lattice sites.

Since the impurity spin acts like a local magnetic field 
the electronic spin density will align along the local field. 
For small values of the coupling there is a negative spin density around the impurity site.
At the impurity site it is positive, as expected. For larger couplings 
the spin density in the vicinity of the impurity site is 
positive. At small couplings the many-body system screens the effect induced by the
impurity inducing fluctuations that compensate the effect of the local field in a way
that the overall magnetization vanishes. However, for strong enough coupling the many-body
system becomes magnetized in a discontinuous fashion.
One interpretation is that if $J$ is strong enough the impurity breaks a Cooper
pair and captures one of the electrons, leaving the other electron unpaired, and
thus the overall electronic system becomes polarized.
The impurity induces a pair of bound states inside the superconducting energy gap, one at
positive energy (with respect to the chemical potential), and another at a symmetric negative
energy. Even though the spectrum is symmetric the spectral weights of the two energy levels
are not the same and their spin content is also distinct. 
The analysis of the local density of states (LDOS) \cite{morr2,first}
shows that for small coupling the lowest positive energy level
has only a contribution from spin $\uparrow$ and the first level
with negative energy (symmetric to the other level) has only contribution from
spin $\downarrow$. The magnitude of the spectral weight at the impurity site is
different for the two states. 
Considering now the second level located in the continuum, both at positive and negative
energies, there is a mixture of both spin components, even though
in the case of the state at positive energy the magnitude of the peak is larger for
the case of spin $\uparrow$ than for the case of spin $\downarrow$ and in the
case of the state in the continuum at negative energy the relative magnitudes of the
two spin components are reversed. 
Considering a higher value for the coupling one finds that the levels inside the gap
approach the Fermi level. There is a critical value of the coupling for which the
two levels cross in a discontinuous way such that it coincides with the emergence of
a finite overall magnetization.
After the level crossing has occurred, the nature of the states changes. The positive energy bound state
has now only a contribution from the spin $\downarrow$ component and vice-versa,
the first negative energy state has only contribution from the spin component $\uparrow$.
As the level crossing occurred the spin content has changed. 
On the other hand, 
in the first state in the continuum, where the two spin components contribute, 
the magnitude of the $\uparrow$ component is now much larger than the $\downarrow$ component
while for a smaller value of the coupling the magnitudes were of similar size. Also, the $\downarrow$
component of the second state of negative energy is now much smaller than the $\uparrow$
component. The second state has to compensate for the spin flip of the
lowest state by increasing the weight of the spin component aligned with the
external impurity spin.

The diagonalization of this Hamiltonian is performed using the Bogoliubov-Valatin transformation in the form
\bea
c_{i\uparrow} &=& \sum_n \left[ u_n(i,\uparrow) \gamma_n - v_n^*(i,\uparrow) \gamma_n^{\dagger}
\right] \nonumber \\
c_{i\downarrow} &=& \sum_n \left[ u_n(i,\downarrow) \gamma_n + v_n^*(i,\downarrow)
\gamma_n^{\dagger} \right]
\eea
Here $n$ is a complete set of states, $u_n$ and $v_n$ are related to the eigenfunctions of Hamiltonian (2),
and the new fermionic operators $\gamma_n$ are the quasiparticle operators. These are chosen
such that in terms of the new operators
\be
H = E_g + \sum_n \epsilon_n \gamma_n^{\dagger} \gamma_n
\ee
where $E_g$ is the ground state energy and $\epsilon_n$ are the excitation
energies. As a consequence
\bea
\left[H, \gamma_n \right] &=& -\epsilon_n \gamma_n \nonumber \\
\left[H, \gamma_n^{\dagger} \right] &=& \epsilon_n \gamma_n^{\dagger}
\eea
The coefficients $u_n(i,\sigma )$, $v_n(n,\sigma )$ can be obtained solving
the Bogoliubov-de Gennes (BdG) equations \cite{degennes}.
Defining the vector
\[ \psi_n(i) = \left( \begin{array}{c}
u_n(i,\uparrow) \\
v_n(i,\downarrow) \\
u_n(i,\downarrow) \\
v_n(i,\uparrow) \end{array} \right)
\]
the BdG equations can be written as
\be {\cal H} \psi_n = \epsilon_n \psi_n
\ee
where the matrix ${\cal H}$ at site $i$ is given by
\[ {\cal H} = \left( \begin{array}{cccc}
-h
-e_F -J_{l_c} \sin \varphi & \Delta_i &
-J_{l_c} \cos \varphi & 0
 \\
\Delta_i^* & h
+e_F -J_{l_c} \sin \varphi & 0 &
-J_{l_c} \cos \varphi
 \\
-J_{l_c} \cos \varphi & 0 & -h
 -e_F +J_{l_c}
 \sin \varphi
 & \Delta_i \\
0 & -J_{l_c} \cos \varphi & \Delta_i^* & h +e_F +
J_{l_c} \sin \varphi
 \end{array} \right)
\]
\noindent where $h=t\hat{s}_{\delta}$ with 
$\hat{s}_{\delta} f(i)=f(i+\delta)$.
The solution of these equations gives both the energy eigenvalues and eigenstates.
The problem involves the diagonalization of a $(4N)\times(4N)$ matrix.
The solution of the BdG equations
is performed self-consistently imposing at each iteration that
\be
\Delta_i = \frac{g}{2} \left[ <c_{i \uparrow} c_{i\downarrow}> -
<c_{i \downarrow} c_{i\uparrow}> \right]
\ee
where $g$ is the effective attractive interaction between the electrons.
Using the canonical transformation this can be written as
\bea
\Delta_i &=& -g \sum_{n,0<\epsilon_n<\hbar \omega_D} \{ f_n \left( u_n(i,\uparrow) v_n^*(i,\downarrow) +
u_n(i,\downarrow) v_n^*(i,\uparrow) \right) \nonumber \\
&-& \frac{1}{2} \left[ u_n(i,\uparrow) v_n^*(i,\downarrow) +
u_n(i,\downarrow) v_n^*(i,\uparrow) \right] \}
\eea
where $\omega_D$ is the Debye frequency,
and $f_n$ is the Fermi function defined as usual like
\[
f_n=\frac{1}{e^{\epsilon_n/T}+1}
\]
where $T$ is the temperature.
We note that the Bogoliubov-de Gennes equations are invariant under the substitutions
$\epsilon_n \rightarrow -\epsilon_n$, $u(\uparrow) \rightarrow v(\uparrow)$,
$v(\uparrow) \rightarrow u(\uparrow)$, $v(\downarrow) \rightarrow -u(\downarrow)$,
$u(\downarrow) \rightarrow -v(\downarrow)$.

\begin{figure*}
\includegraphics[width=0.4\textwidth]{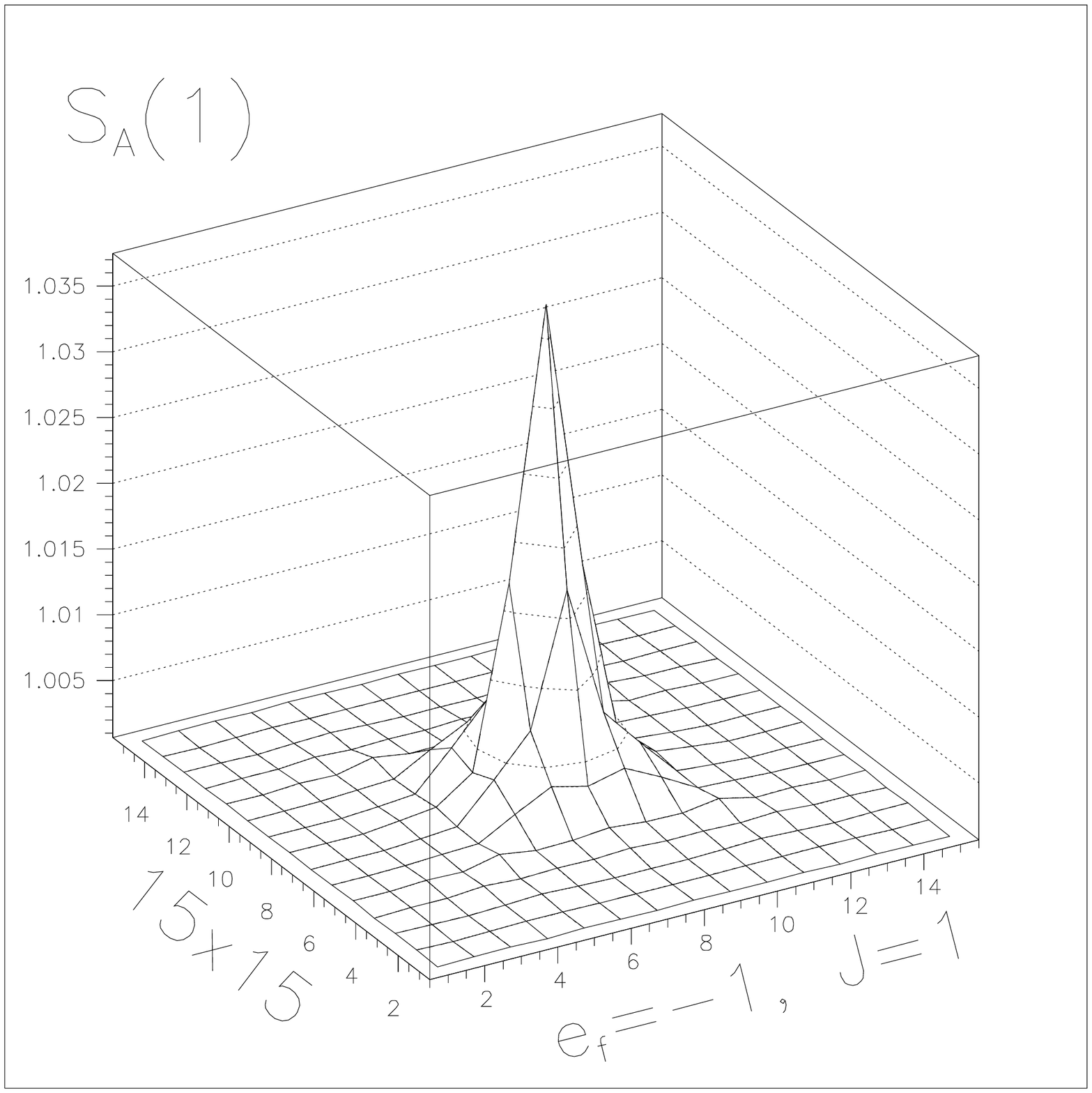}
\includegraphics[width=0.4\textwidth]{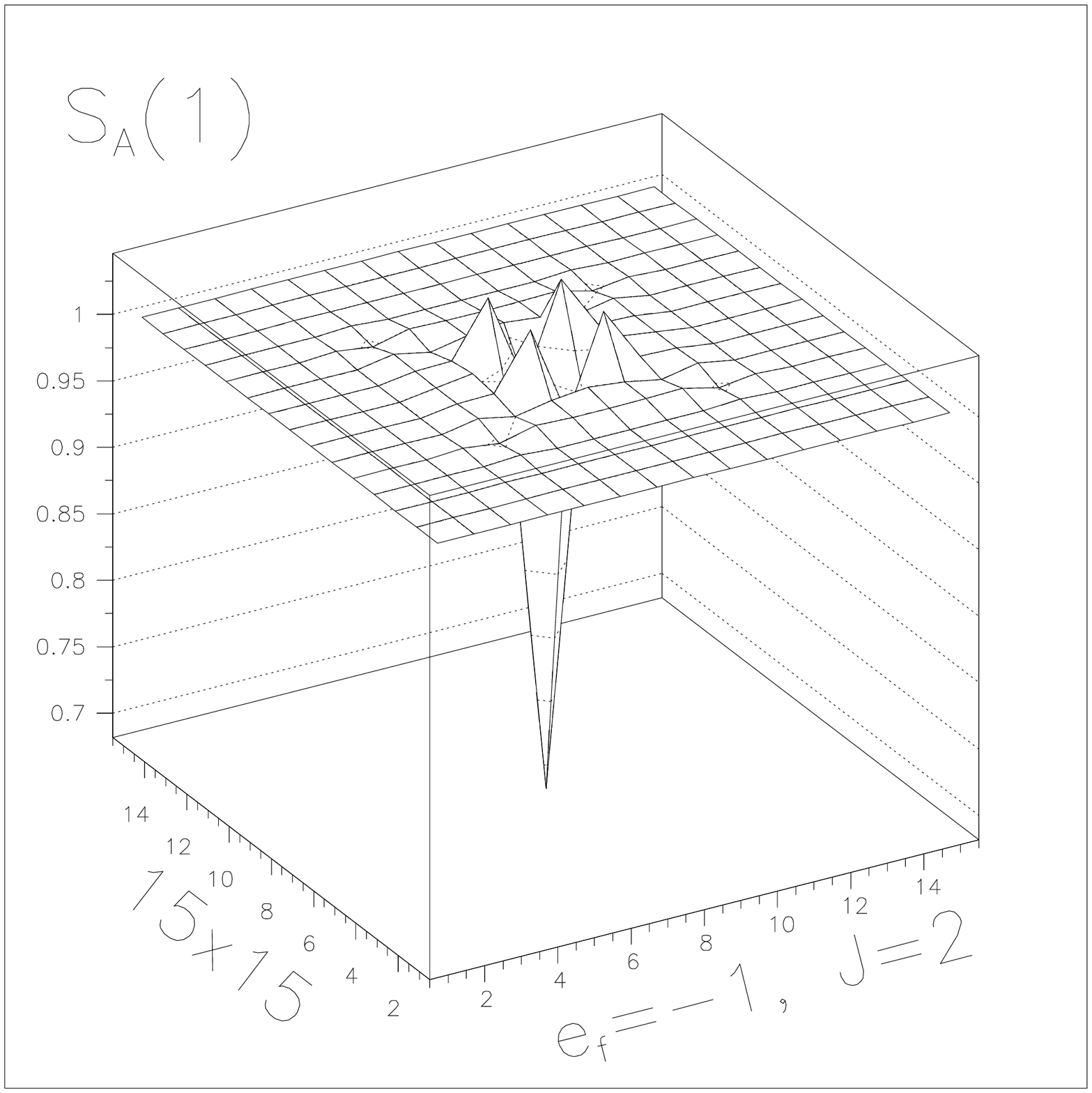}
\includegraphics[width=0.4\textwidth]{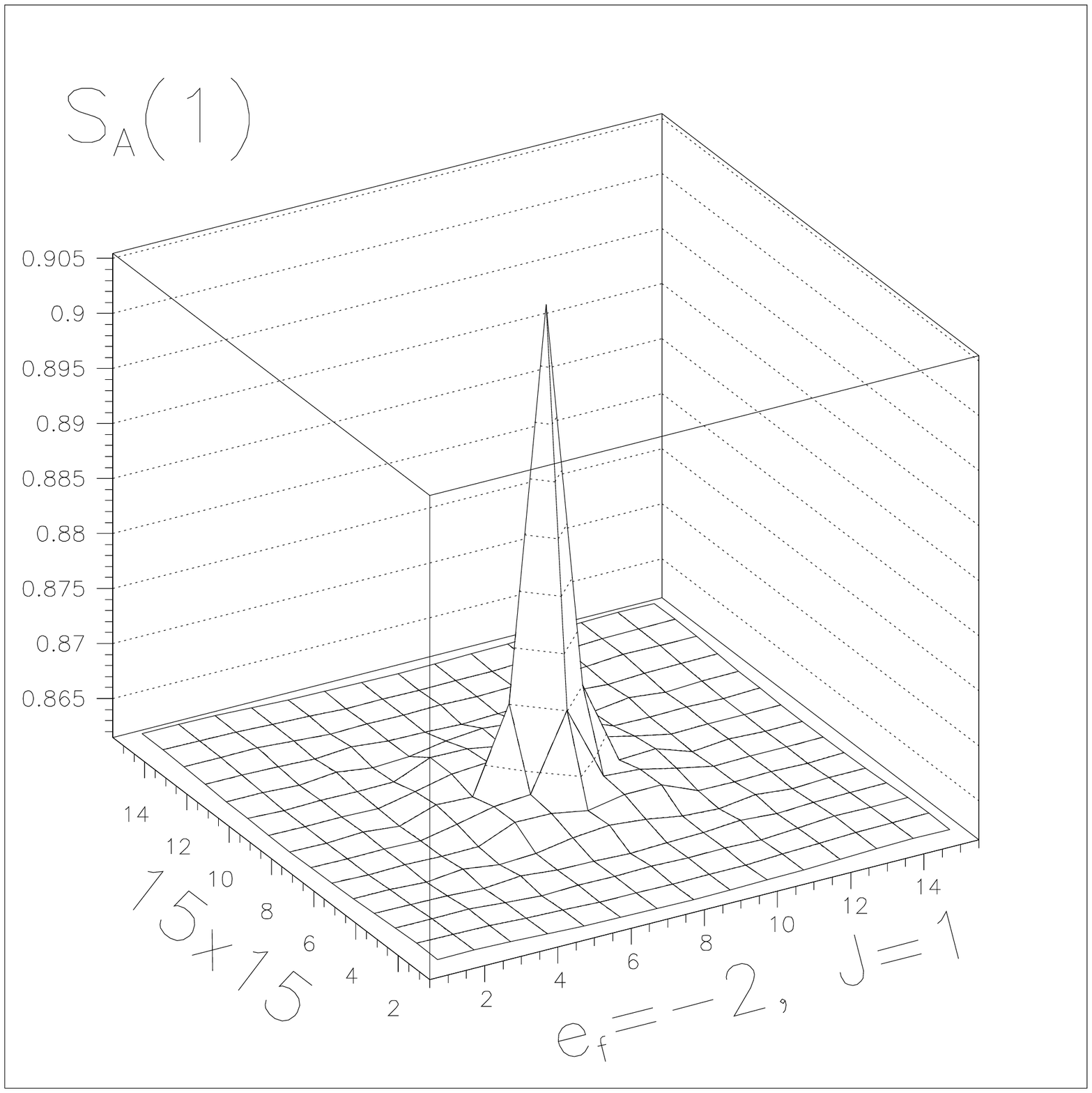}
\includegraphics[width=0.4\textwidth]{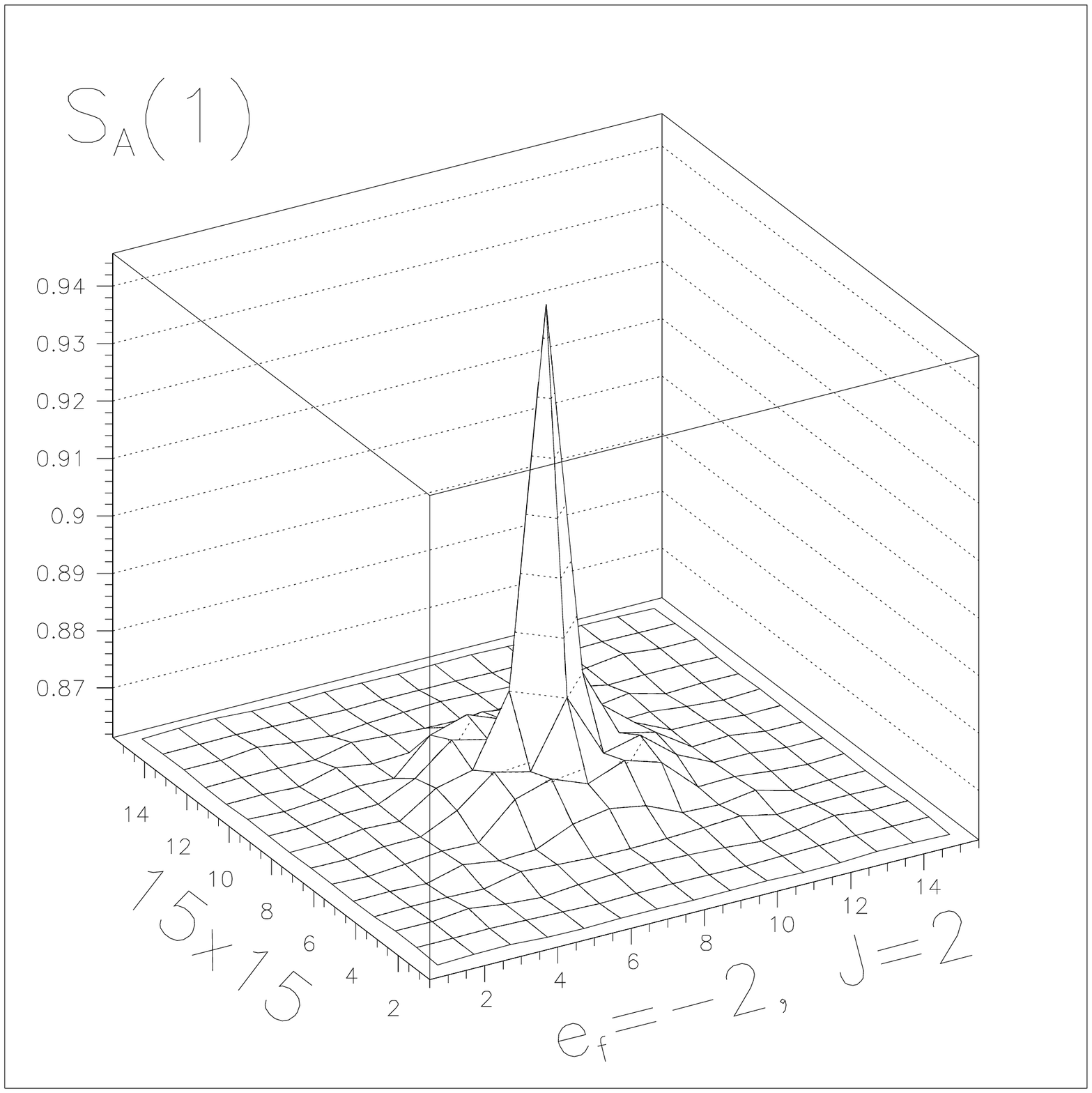}
\caption{\label{fig1}
Single-site von Neumann entropy $S_A(1)$ for $J=1,2$ and for different 
chemical potentials $e_F=-1,-2$. The system size is $15 \times 15$.
Note that in this figure and in the others the entropy is expressed
in the basis $e$.
}
\end{figure*}

\section{von Neumann entropy}

Consider a system at zero temperature so that its density matrix
 respects the ground state. If this is a pure state, as in our problem,
the density matrix is simply
\be
\rho = |\Phi \rangle \langle \Phi |
\ee
where $|\Phi\rangle$ is the ground state, which in our case is the BCS ground state wave-function.
Consider now that one can divide the system into two parts A and B, and define
\be
\rho_A = Tr_B \rho .
\ee
This quantity characterizes the reduced density matrix of the subsystem A, having integrated out
the degrees of freedom of the rest of the system, B. However, the information on the correlations
intrinsic to part B and the correlations between the two subsystems is implicit.
We can therefore define the so-called von Neumann entropy as the information contained on part A
of the system in the usual way
\be
S_A= -Tr \rho_A \ln_2 \rho_A
\ee
This quantity is one possible measure of the entanglement between subsystem A and the rest of the
system.

The subsystem A may have different extensions. The simplest case corresponds to select A as a
single site and B the remaining $N-1$ sites. In a translationally invariant system, the entropy
$S_A$ will be the same for every site, but in a non-homogeneous system, as in our problem,
the entropy is site dependent, since the impurity breaks the translational invariance.
As we will see below the von Neumann entropy is site dependent particularly in the vicinity
of the impurity. Also, we will see that it depends on the spin coupling, particularly as
the quantum phase transition is crossed. It is also interesting to look at the two-site
entanglement where the subsystem A is now composed of two sites (arbitrarily selected from
the lattice) and B are the remaining $N-2$ sites. This quantity also shows a behavior that
clearly identifies the phase transition.

In many-body systems a description in terms of wave functions is quite involved and second
quantization is the natural way to perform any calculation. The density matrix is however
easily defined in terms of states. It is also easily defined in terms of its matrix elements
in some basis and this approach is indeed easy to implement using Fock states. It is easy therefore
to show that the matrix elements of the density matrix are simply defined in terms of correlation
functions of the whole system. For instance, in the case of the single-site entanglement it can be shown
that, using a basis of local states like $|0\rangle, |\uparrow, \downarrow \rangle,
|\uparrow \rangle, |\downarrow \rangle $, which denote the four possible
 states --- unoccupied, double occupied, singly occupied with an electron
 with spin $\uparrow$ and singly occupied with an electron with spin
 $\downarrow$, respectively --- the density matrix can be defined as
\[ {\cal \rho}_A = \left( \begin{array}{cccc}
 \langle (1-n_{\uparrow}) (1-n_{\downarrow}) \rangle & \langle c_{\uparrow}^{\dagger} c_{\downarrow}^{\dagger}
\rangle & 0 & 0
 \\
\langle c_{\downarrow} c_{\uparrow} \rangle & \langle n_{\uparrow} n_{\downarrow} \rangle & 0 & 0
 \\
0 & 0 & \langle n_{\uparrow} (1-n_{\downarrow}) \rangle & \langle c_{\downarrow}^{\dagger} c_{\uparrow} \rangle 
 \\
0 & 0 & \langle c_{\uparrow}^{\dagger} c_{\downarrow} \rangle & \langle (1-n_{\uparrow}) n_{\downarrow}\rangle
 \end{array} \right)
\]
The spin and charge parts decouple. The spin part couples the two spin orientations and the charge
part couples the empty and doubly occupied cases. The diagonal terms of the matrix describe the
number of empty sites, number of doubly occupied sites, the number of $\uparrow$ spin sites and
the number of $\downarrow$ spin sites, respectively. This matrix is easily diagonalized and the
von Neumann entropy obtained straightforwardly. The sum of the diagonal terms is equal to $1$ due to normalization.

The correlation functions are easily solved using the representation of the electron
operators in terms of the BdG quasiparticle operators.
Specifically, we can write the single-site density matrix as
\be
\rho = \sum_{n,m} |n\rangle \rho_{nm} \langle m|
\ee
where $|n\rangle,|m \rangle $ are the four states described above.
Consider now an operator $O$ defined in terms of
creation and annihilation operators of the electrons, and evaluate
\be
Tr \{ O \rho \} = \sum_{n,m} \langle n| O |m \rangle \rho_{m,n} .
\ee
To determine the matrix element $\rho_{nm}$ of the density matrix it
is enough, by inspection, to find which operator $O$ is such that only
the matrix element
$\langle n| O |m \rangle $
is non-zero. This will tell us which correlation function
gives which matrix element of the density matrix. 
Alternatively, one can express the operator $|m \rangle \langle n|$ in terms of
the creation and annihilation operators.
The results for $\rho_A$ are in the equation referred above.

These matrix elements are defined in the subspace of the states of one site.
However, since we have integrated out all the other sites, we may now
replace the matrix elements by the matrix elements over the entire
system (direct product of the states of the remaining $N-1$ sites).

The case of the two-site entanglement is much more involved. One introduces two sites, $i$ and $j$,
and uses the same basis states for each site. The basis states are given by the direct product of these
two basis sets and the reduced density matrix is now a $16 \times 16$ matrix.
It is convenient to organize the basis states into even-even and odd-odd states in the number
of electrons in the sites $i$ and $j$, respectively. These two sets of states decouple and we
are reduced to the diagonalization of two $8 \times 8$ matrices.
The calculation of the various correlation functions involved in these two
matrices would be rather lengthy if performed by hand.
Note that one of the correlation functions involves the double occupancies
at the two sites.
This involves a product of eight electron operators, and each electron
operator is given by a sum of two quasiparticle operators.
Also one should pay special attention to the anti-commutators of the
electron and quasiparticle operators.
It is therefore advisable to calculate all of these correlation functions
in an automatic way, using a symbolic algebra package, and we did so
with the help of the computer program FORM\cite{FORM}.

\subsection{Single-site von Neumann entropy}

Let us consider the results for the single-site entropy. We will present results for a system of size 
$15 \times 15$. As we have shown previously \cite{first} the nature of the phase transition is not
affected by changing either the system size or the boundary conditions (ie they can be either open or periodic).
As shown previously, if the system is large enough the states induced by the impurity are localized and therefore
not affected by the boundary. Also, since in this problem we are
dealing with a first order quantum phase transition, not a second order one, there are
no long range critical fluctuations that would imply a more careful scaling analysis. So we will
limit ourselves to an adequate system size that illustrates the sensitivity of the entanglement
measures to the ground state phase transition. 

In Fig. \ref{fig1} we show the results for the single-site entropy for different chemical
potentials. In the case $e_F=-1$ the quantum phase transition is signaled
by the change in the entropy near the impurity location. Below the QPT there is a maximum
and as the QPT occurs the entropy develops a local minimum as the system gets correlated. 
Before the QPT is reached the impurity leads to a local increase of the entropy since it acts against the order
due to the superconducting phase. At the QPT the physical
picture is that the impurity locks an electron, which causes a local decrease in entropy.
In the case $e_F=-2$ a value of $2$ for the coupling $J$ is not strong enough to cross the phase transition.

\begin{figure}
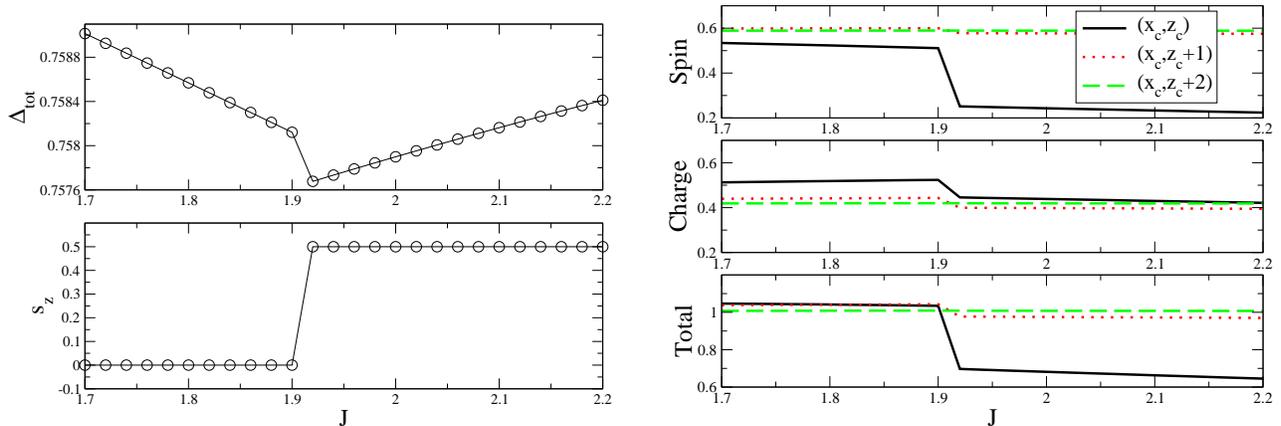

\includegraphics[width=0.45\textwidth]{Fig2a}
\hspace{0.5cm}
\includegraphics[width=0.45\textwidth]{Fig2b}
\caption{\label{fig2}
a) Total gap function and spin magnetization and b) spin, charge and total contributions to the
single-site entropy as a function of couping. The curve in black corresponds to the impurity location, $l_c$,
and the other two curves to two points distant by one lattice spacing and two lattice spacings.
 Notice the discontinuities signaling the QPT.
}
\end{figure}

\begin{figure}[h]
\includegraphics[width=0.3\textwidth]{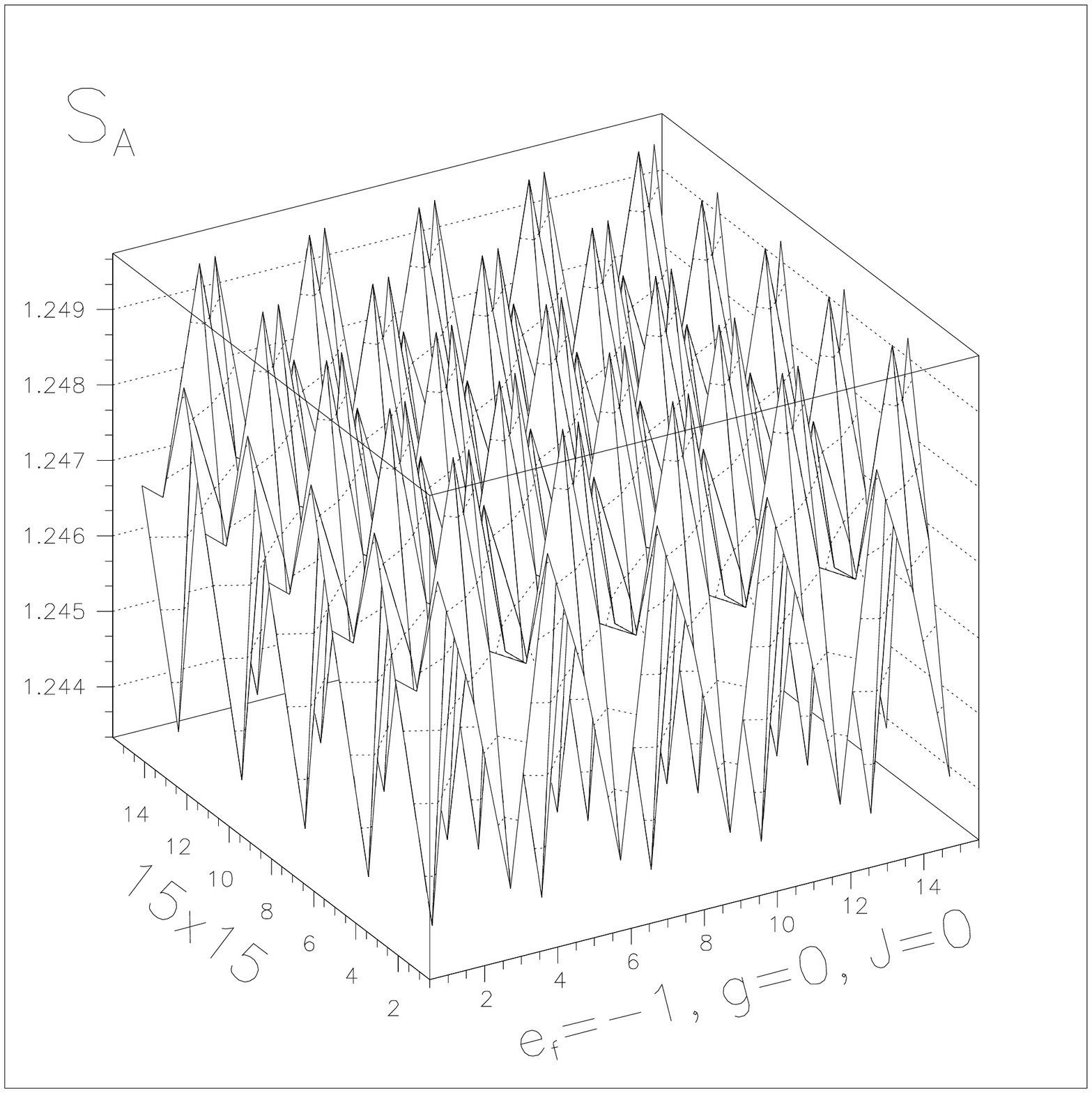}
\includegraphics[width=0.3\textwidth]{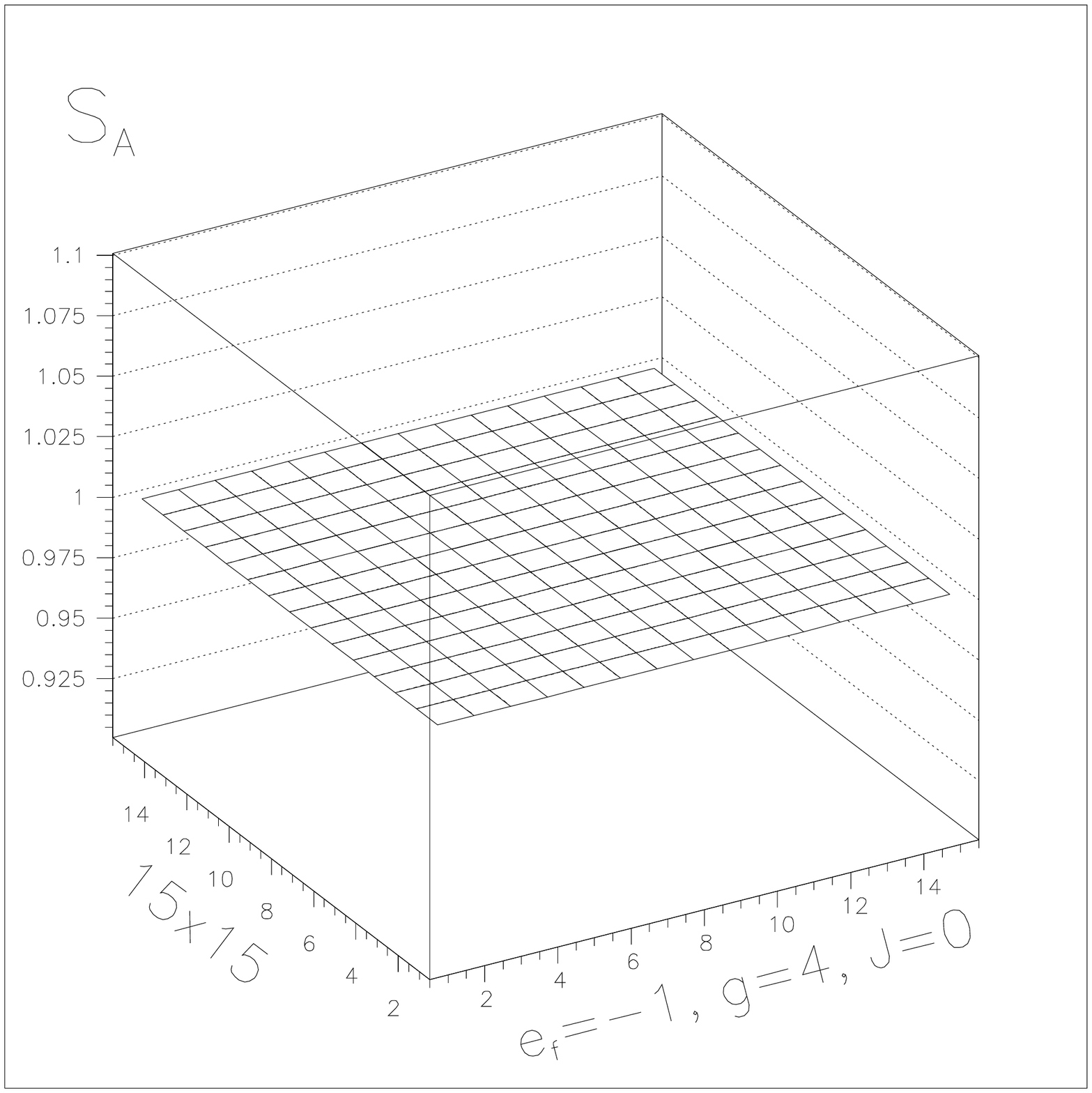}
\includegraphics[width=0.3\textwidth]{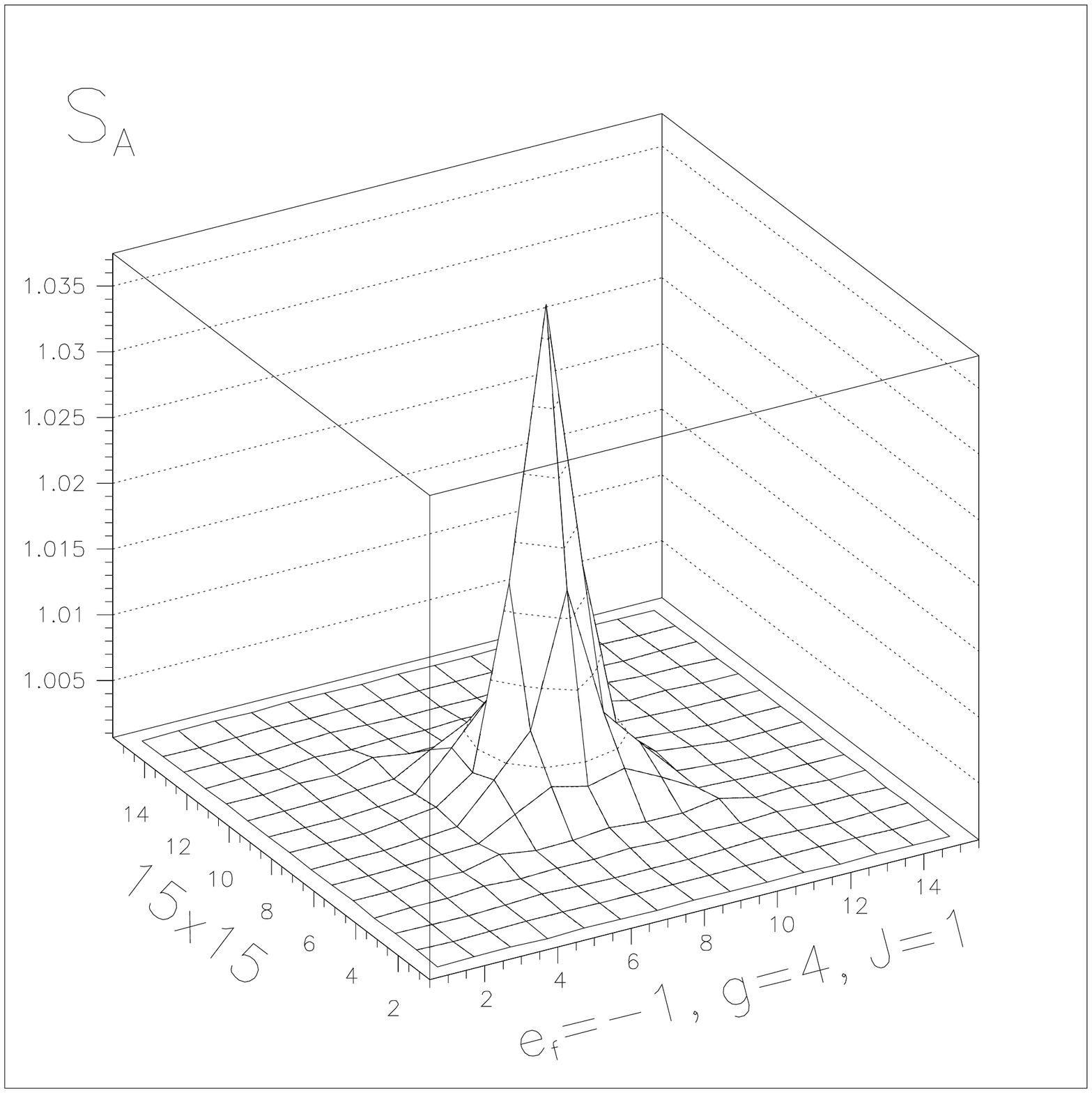}
\caption{\label{figclarify}
Single-site entropy $S_A(1)$ for a tight-binding system ($g=0,J=0$), 
a superconductor ($g \neq 0, J=0$) and a superconductor
with a magnetic impurity ($g \neq 0, J=1$).
}
\end{figure}

In Fig. \ref{fig2} we show the QPT signatures in terms of the total gap function and the total
magnetization, and in terms of the single-site entropy. Notice that as the coupling increases
the total magnetization becomes finite at the QPT, discontinuously, which is indicative of a first
order phase transition. Note the small change on the average value of the gap function as the coupling
changes. However, the trend is clearly altered as the QPT occurs.
 As the figure also shows, the largest entropy change is related to the
spin degrees of freedom, as one might expect due to the spin interaction. However, as the
trace of the reduced density matrix is normalized, the charge part is also discontinuous
at the phase transition. 
It is also interesting to note that, as expected, at the impurity site,
and for values of $J$ higher than the critical value, the spin contribution
to the entropy is smaller than the charge contribution.

In order to better clarify the meaning of the von Neumann entropy, in Fig. \ref{figclarify} we present a comparison of
the single-site entanglement entropy for three different cases, namely i) a free (tight-binding) system where there is
no superconductivity and no magnetic impurity, ii) a superconductor with no
magnetic impurity, and iii) a superconductor with an impurity. 
The entropy is basically uniform in the
free case and in the superconductor. The value in the free case is larger, 
as expected due to the ordered nature of the superconductor. 
Introducing the impurity changes the entropy in the vicinity of the
impurity but does not alter the background value.

\subsection{Two-site von Neumann entropy}

Consider now the two-site entanglement. The von Neumann entropy now measures the entanglement between
two sites and the remaining $N-2$ sites. 

\begin{figure*}
\includegraphics[width=0.4\textwidth]{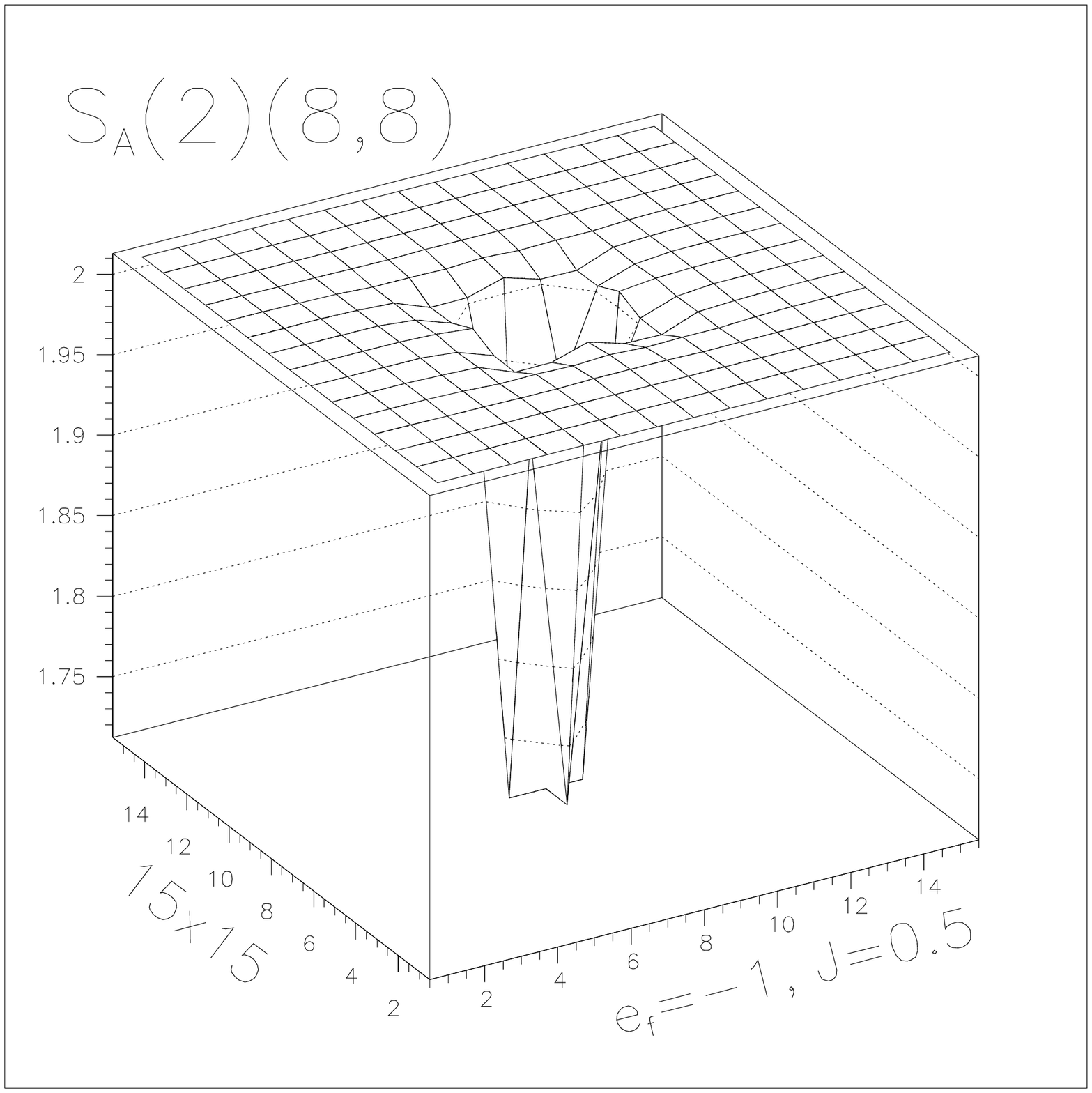}
\includegraphics[width=0.4\textwidth]{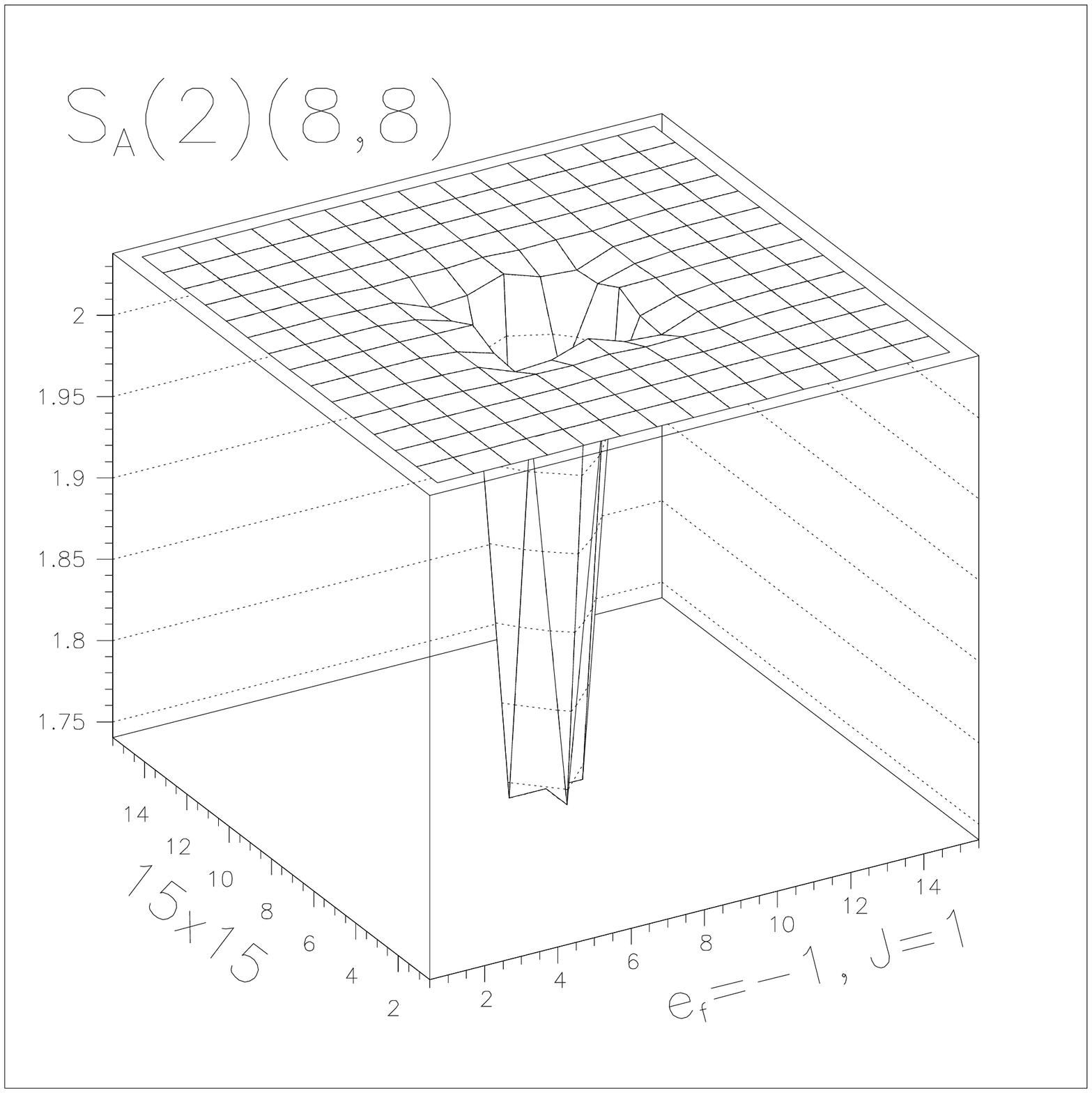}
\includegraphics[width=0.4\textwidth]{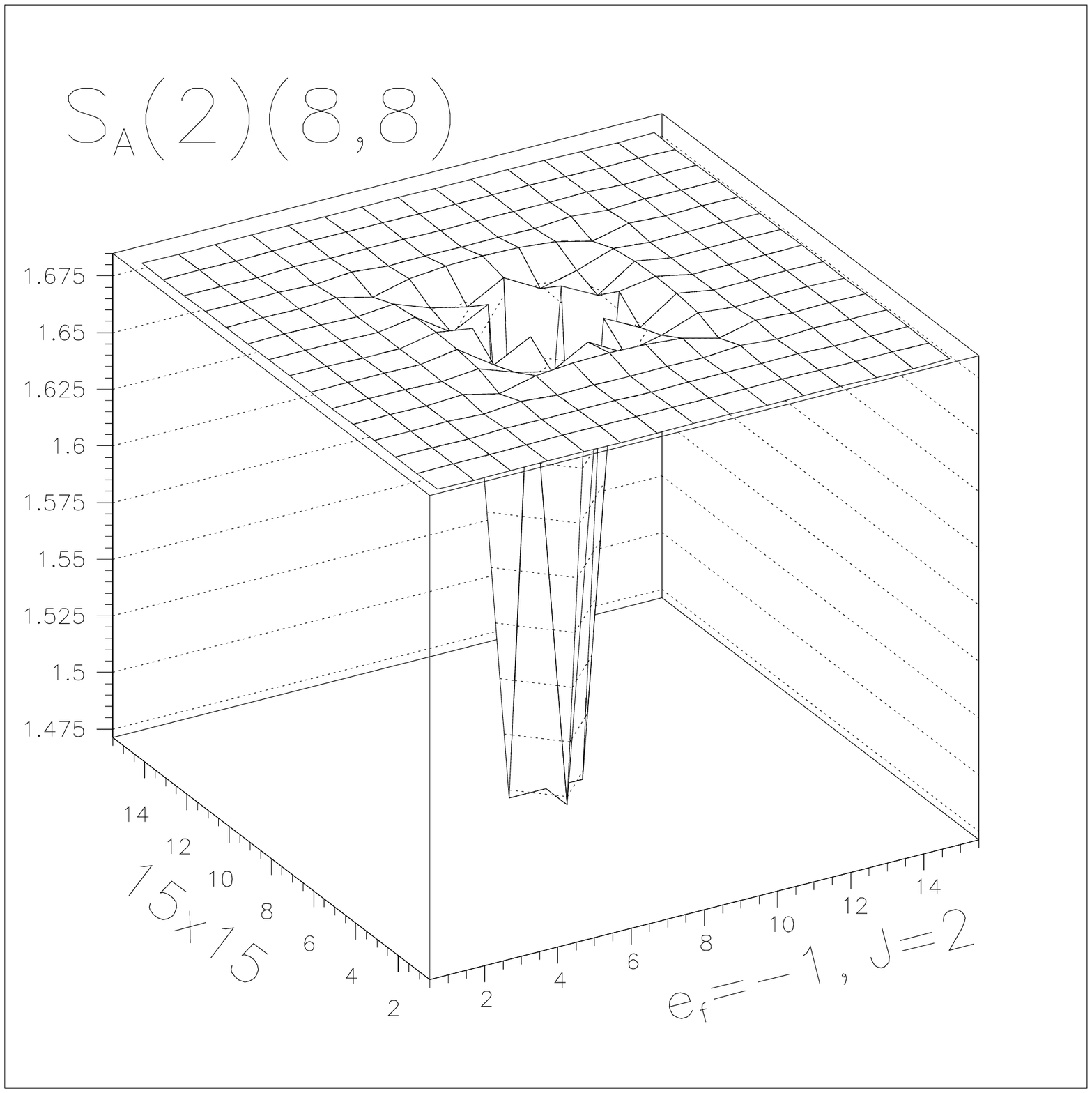}
\includegraphics[width=0.4\textwidth]{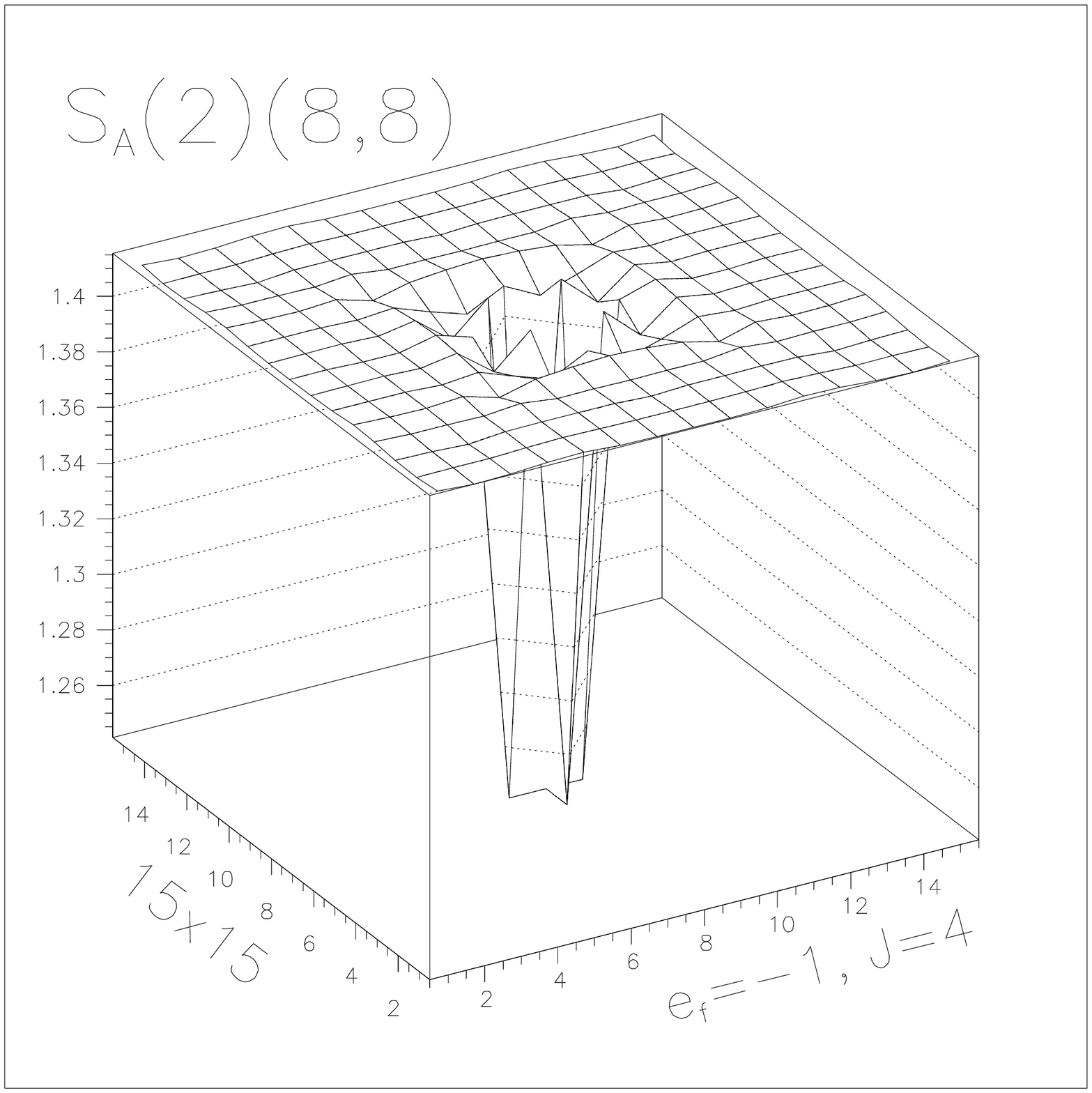}
\caption{\label{fig3}
Two-site entropy $S_A(2)$ where $i=(x_c,z_c)=(8,8)$ is at the 
impurity location for a system with $15\times 15$ sites
and $j$ is at any other location, for $J=0.5,1,2,4$.
}
\end{figure*}

In Fig. \ref{fig3} we present the two-site entanglement entropy choosing the spin impurity as a fixed site
and letting the second site be any other site of the system. Notice the decrease
in the entanglement entropy when the second site is in the vicinity of the fixed site. 
This is natural since one expects close neighbors to be more strongly correlated than further distanced sites,
and therefore the entropy should be lower in the first case.
Also, as the QPT transition is crossed, even though the structure of the two-site entanglement
entropy is not strongly affected one may observe that the background value steadily decreases. This is consistent
with the decrease of the single-site entanglement entropy in the vicinity of the impurity.

\begin{figure*}
\includegraphics[width=0.4\textwidth]{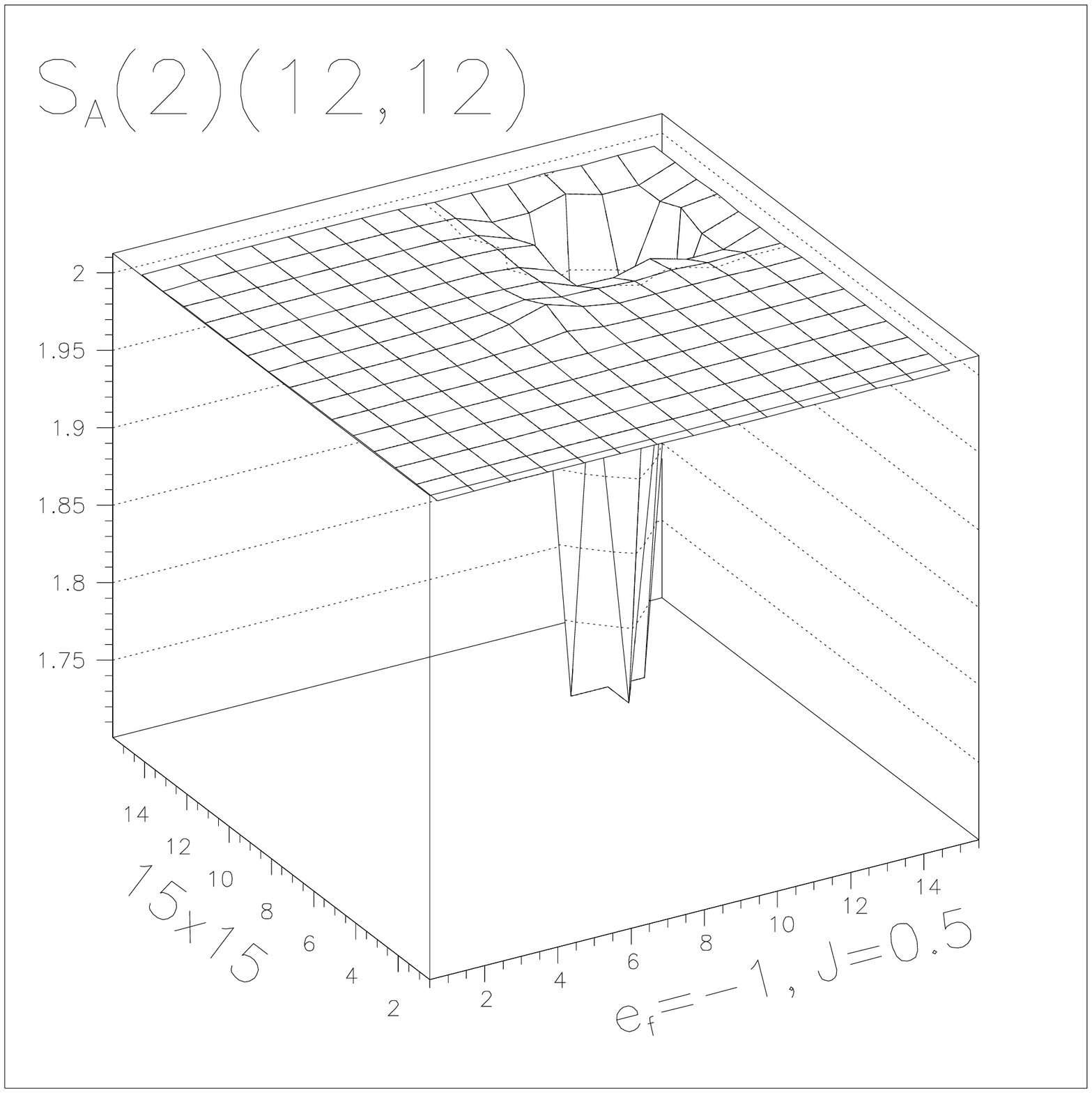}
\includegraphics[width=0.4\textwidth]{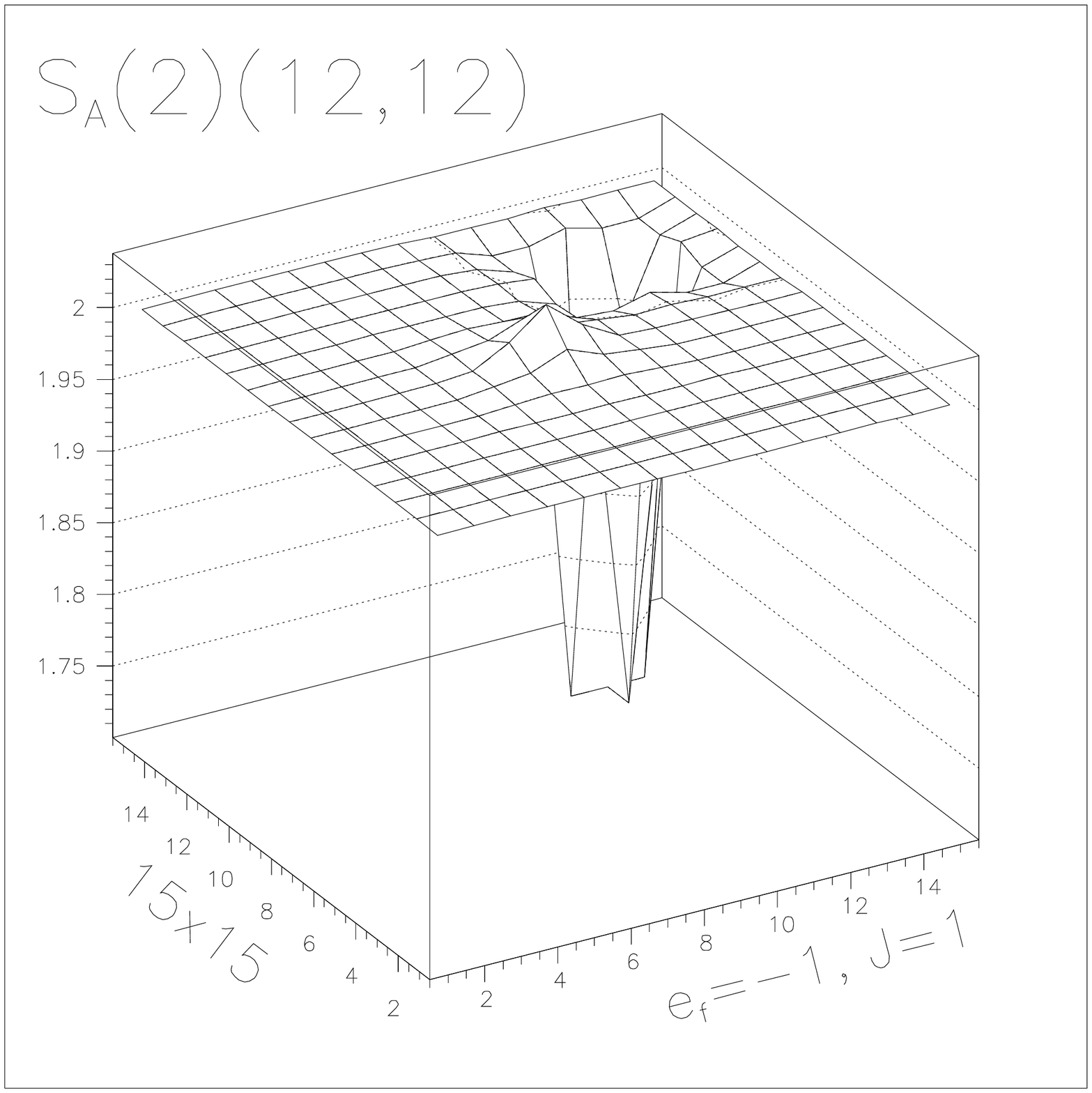}
\includegraphics[width=0.4\textwidth]{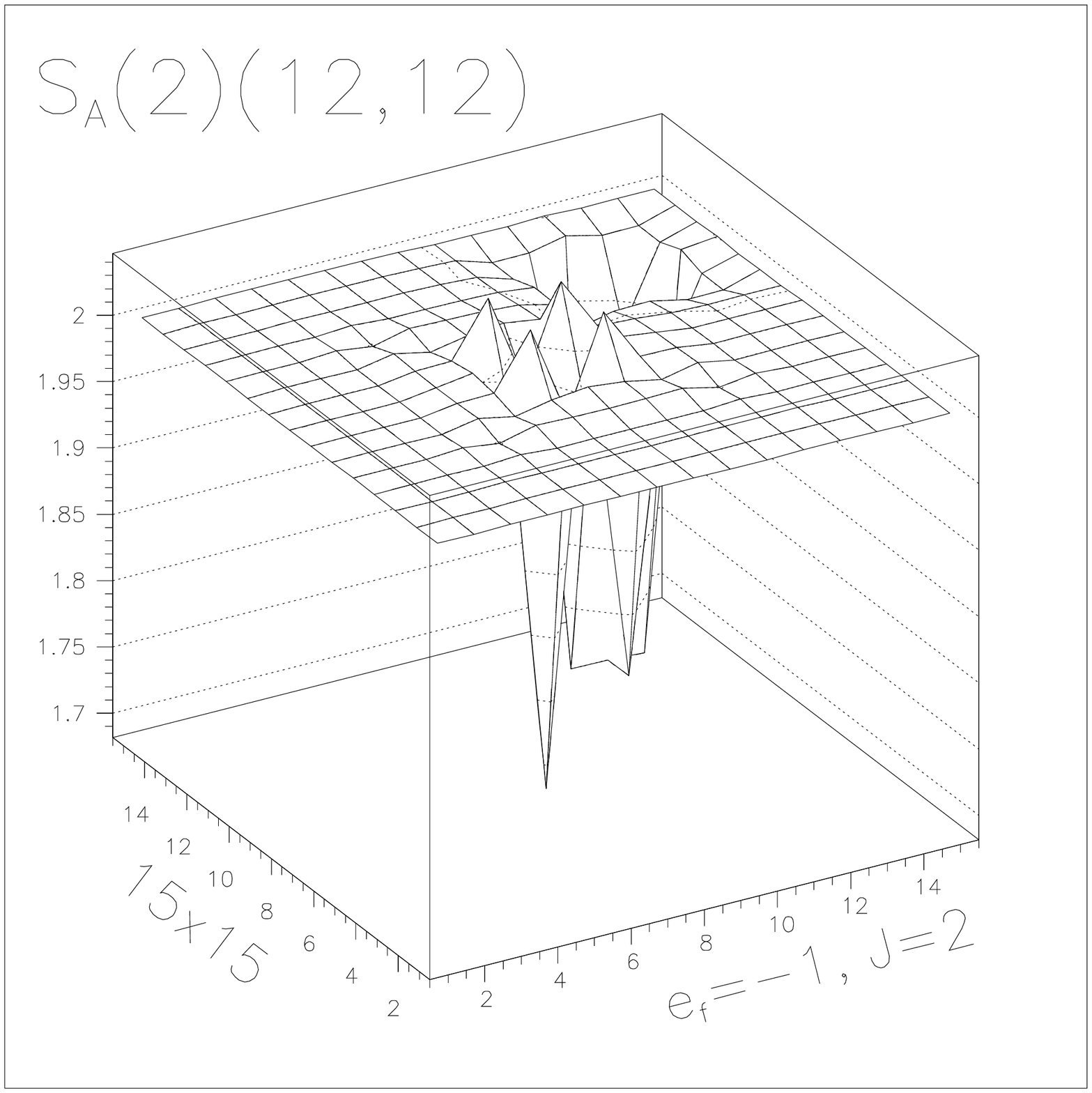}
\includegraphics[width=0.4\textwidth]{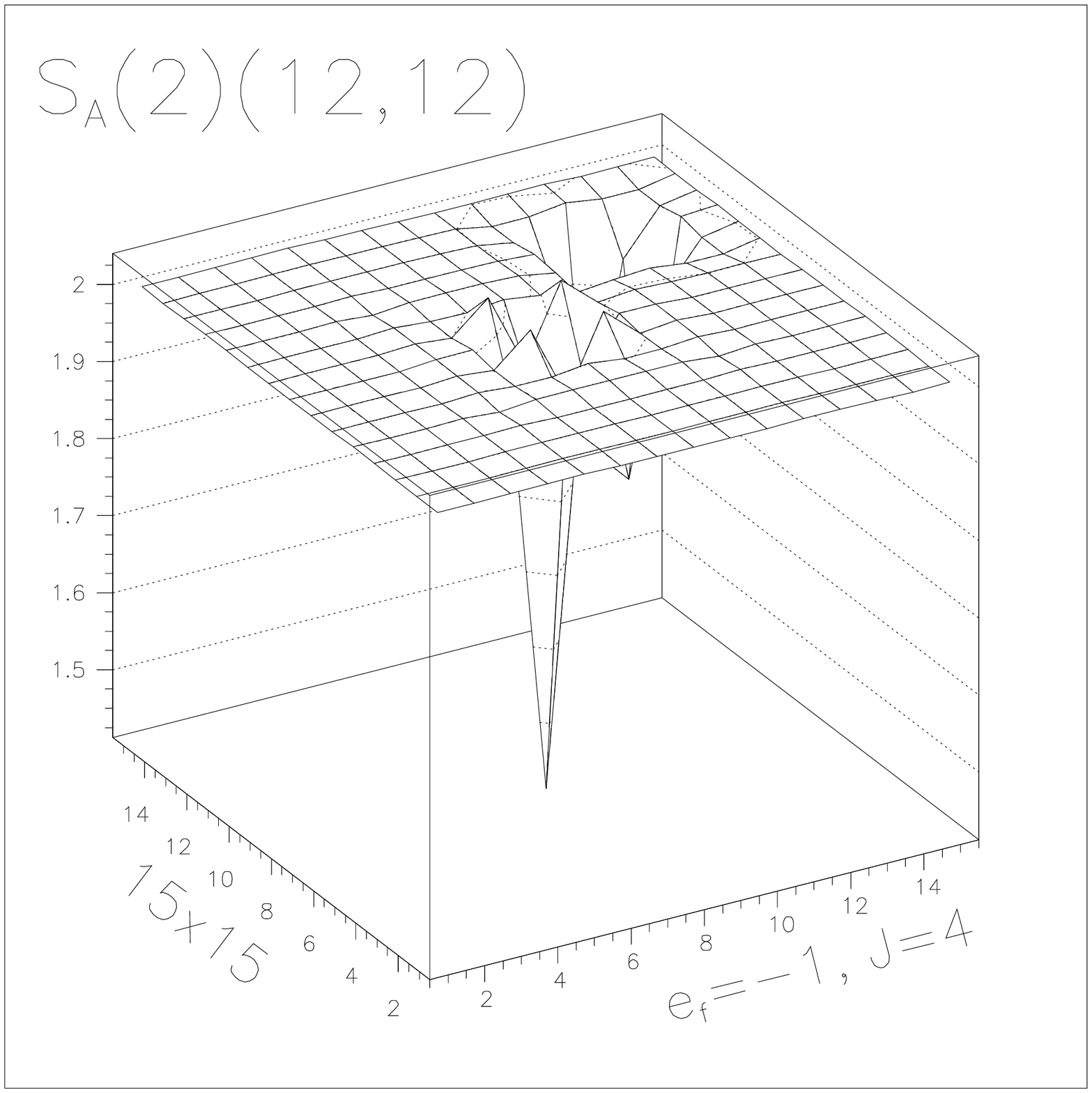}
\caption{\label{fig4}
Two-site entropy $S_A(2)$ where $i=(12,12)$ is a site in the bulk 
and $j$ is at any other location, for $J=0.5,1,2,4$.
}
\end{figure*}

In Fig. \ref{fig4} we present some results for the case where the fixed site is now in the bulk. Notice that
as the coupling increases one can observe two structures: one related to the points in the vicinity
of the bulk site, where the entropy is smaller, and another structure near the impurity site.
Notice that in this case the background value is not affected when the QPT is crossed.

\begin{figure}
\includegraphics[width=0.4\textwidth]{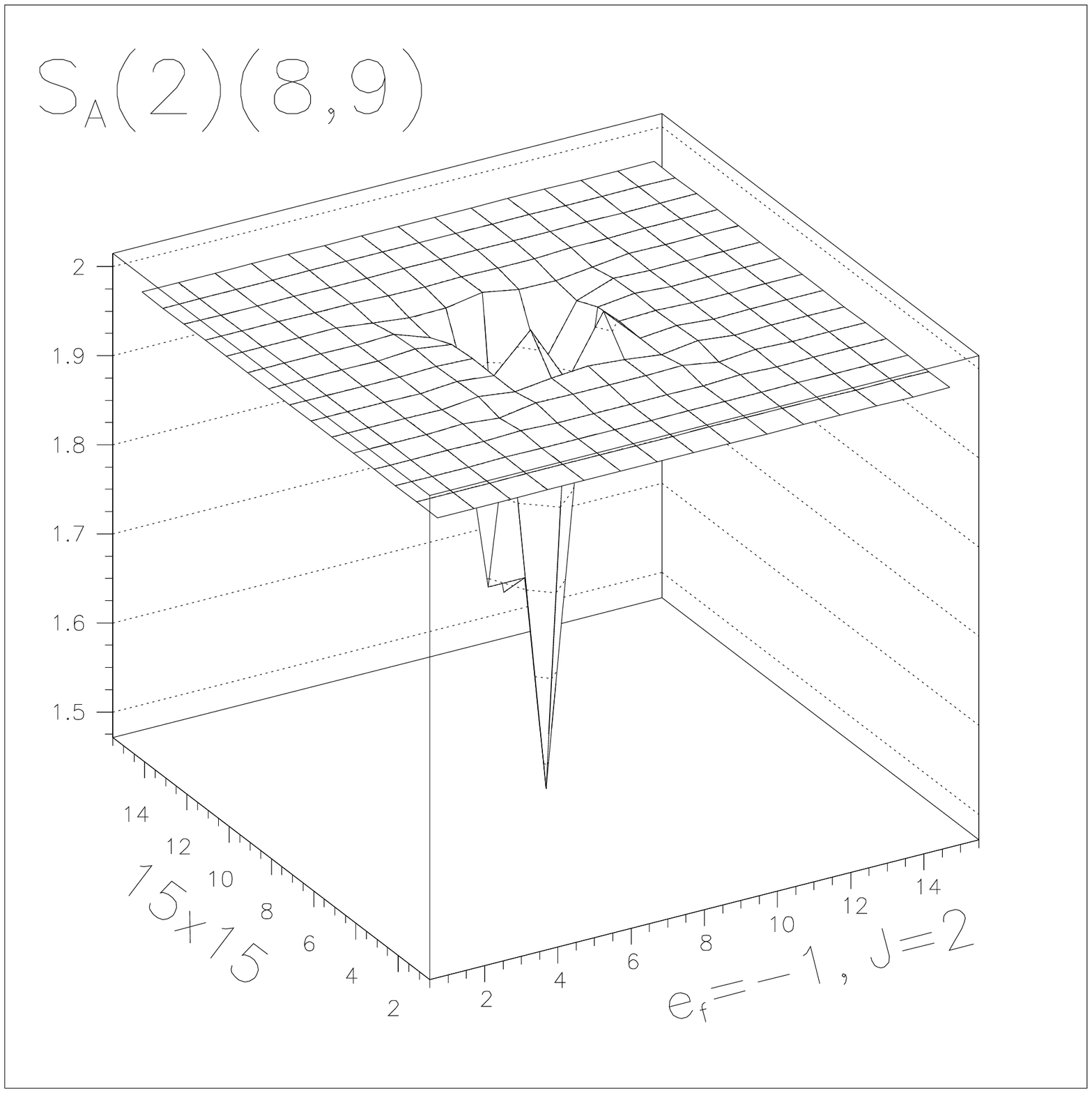}
\hspace{0.5cm}
\includegraphics[width=0.4\textwidth]{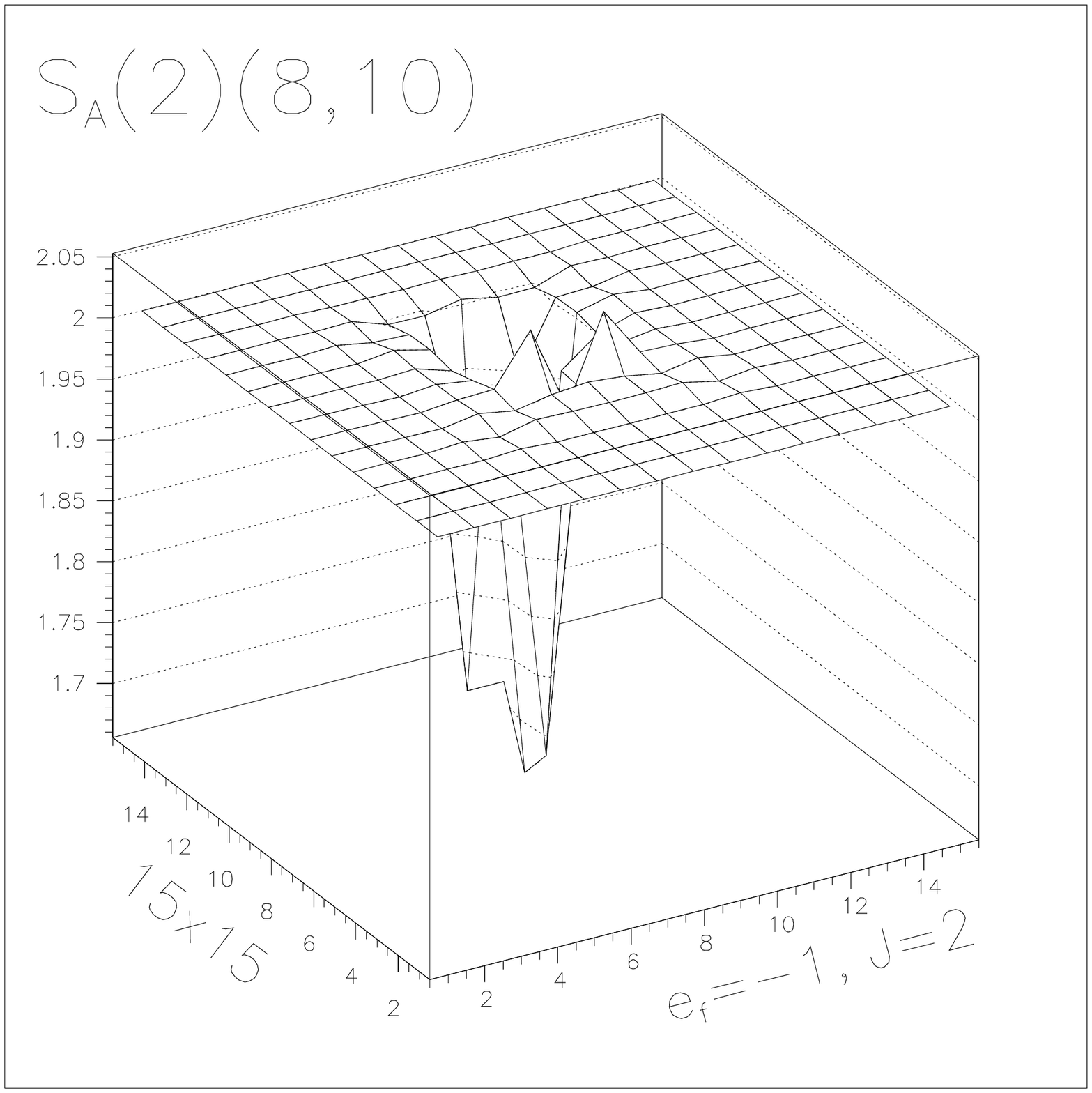}
\caption{\label{figclarify3}
Two-site entropy $S_A(2)$ for $J=2$ where a) one of the points is fixed at 
a nearest-neighbor of the impurity and b) one of the points is fixed
in the next-nearest-neighbor. Note that the background value is different
with respect to the case where one of the points is fixed at the impurity.
}
\end{figure}

We may as well compare the two-site entropy for a free system, a superconductor, and a superconductor with the impurity
inserted. For instance it is interesting to consider the case where one of the points of subsystem B is the 
impurity site.
In the tight-binding case one finds small ripples (as for the single-site entanglement) which are smoothened
in the superconductor. Also, the background value decreases when the system becomes superconductor. 
In all cases there is a depression in the two-site entanglement when the points
$i$ and $j$ are nearby. The effect of the impurity is to narrow somewhat the extent of that depression, which therefore becomes
hardly noticeable. 
In Fig. \ref{figclarify3} we compare the two-site entanglement when one of the sites
is the impurity site and the other any other
as compared to a case where neither includes the impurity site, including a nearest-neighbor. 
The results show that the
effect of the impurity is noticeable since the background value is strongly affected. 

\begin{figure}[b]
\begin{picture}(300,230)
\put(-10,0){\includegraphics[width=0.55\textwidth]{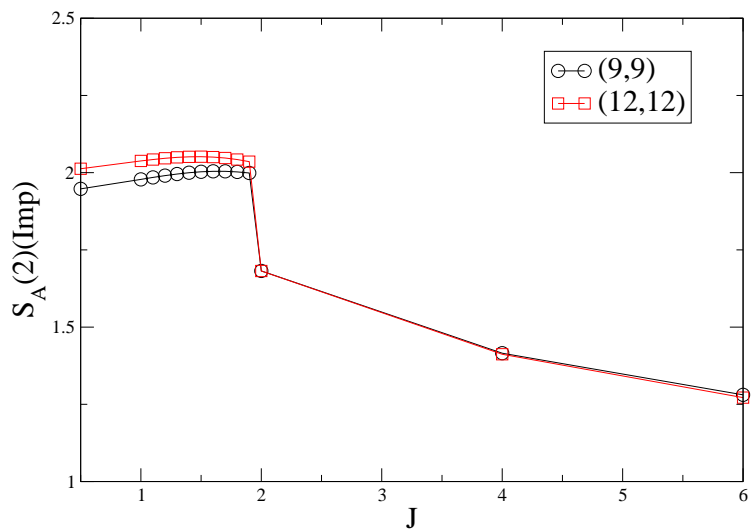}}
\end{picture}
\caption{\label{figsignal}
Two-site entropy $S_A(2)$ as a function of $J$ where one of the points is 
fixed at the impurity location and the other at another site in the bulk.
}
\end{figure}

The QPT can therefore be as well signaled by the two-site entanglement entropy. In Fig. \ref{figsignal}
we show, as a function of the coupling $J$, the two-site entropy in two cases:
 i) between a site in the bulk and any other site in the system,
and ii) between the impurity site and any other site in the system.
 There is a clear regime difference in both cases as the critical point is crossed.


\subsection{Mutual information}

The von Neumann entropy measures the entanglement between the subsystem A and the rest of the system, denoted
subsystem B. However, looking at the entanglement between two sites of the system should provide us with a more detailed view.
We may define the reduced density matrix of a subsystem of two sites, by integrating as before the
degrees of freedom of the remaining ($N-2$ sites). We may now look at the entanglement between the two sites.
In general the states between two sites may be divided into entangled and disentangled states.
A subset of the disentangled states is given by the product states of the two sites. The entanglement
between the two sites of subsystem A may be defined in the following way.
The states are characterized by their density matrices. In a standard notation we will characterize the entangled
states of two sites $1$ and $2$ by a density matrix $\rho_{12}$ (which in our case is a reduced density
matrix). A disentangled state that may be written as a product state is characterized by a density
matrix $\eta_{12}$ and a general disentangled state is
characterized by a density matrix $\sigma_{12}$. The entanglement between the sites $1$ and $2$ is
defined as the ``distance''
\be
E_{12} =\inf\, Tr \, \rho_{12} \left( \log \rho_{12} - \log \sigma_{12} \right)
\ee
The procedure to determine the infimum (minimum) of that trace is in general very complex.
Another quantity that is much easier to calculate is the so-called mutual information where instead
of looking for the ``distance'' over all disentangled states we just look for the minimum over
the product states. This can be shown to lead to
\be
M_{12} = S_A^{(1)}(1) + S_A^{(1)}(2) - S_A^{(2)}(1,2)
\ee
where $S_A^{(1)}$ is a single-site von Neumann entropy and $S_A^{(2)}$ is a two-site von Neumann entropy.
While $E_{12}$ measures the full entanglement between the two sites, the mutual information just
measures the correlation. 

\begin{figure}
\includegraphics[width=0.7\textwidth]{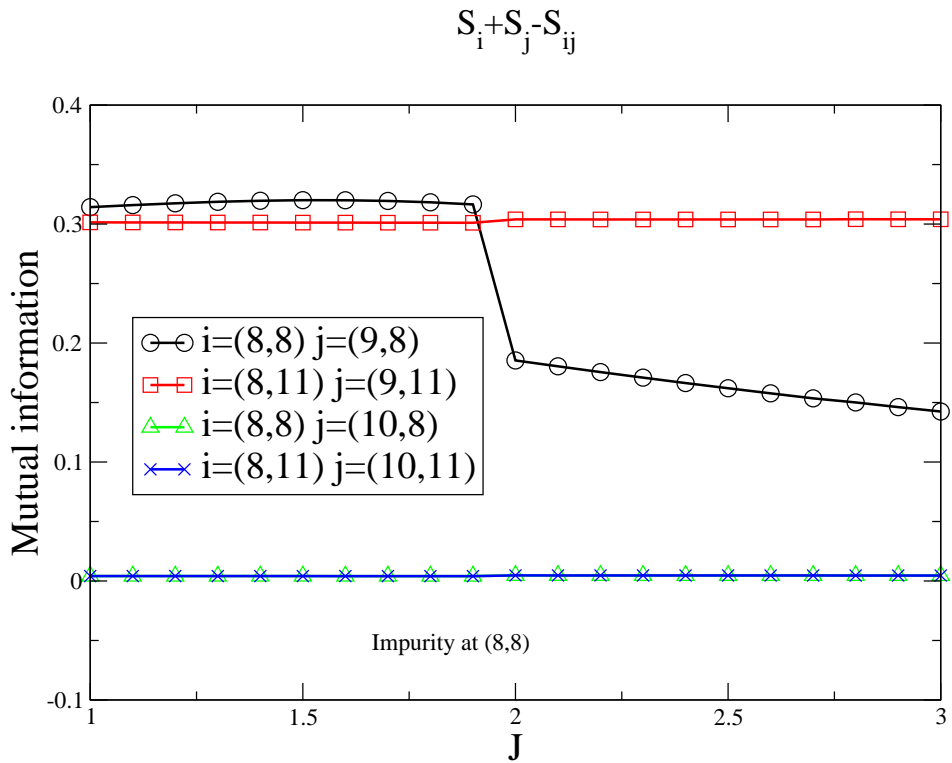}
\caption{\label{mutual}
Mutual information as a function of $J$ where one of the points is fixed at the impurity location $(8,8)$
and the other at two sites in the bulk $(9,8),(10,8)$ and one of the points is at a point in the
bulk $(8,11)$ and two other points a nearest neighbor $(9,11)$ and a second 
nearest neighbor $(10,11)$. 
}
\end{figure}

In Fig. \ref{mutual} we compare the mutual information for different pairs of sites. When the sites
are nearest neighbors the mutual information is finite and is otherwise quite small, even for second
nearest neighbors. The QPT is clearly seen, particularly when one of the sites is the impurity site.

\section{Negativity}

Another quantity that has been proposed to distinguish between entanglement and non-entanglement
is the so-called negativity. One considers two subsystems A and B, and constructs the joint
density matrix of the two subsystems, call it $\rho_{AB}$. Define now the transposed in A density
matrix in the sense that the basis states of part A are transposed while the basis
states of part B remain the same. Looking at the eigenvalues of this new transposed density
matrix one finds that it may have negative eigenvalues, even though the original density matrix never does.
The existence of negative eigenvalues means that the system is necessarily
entangled and therefore one possible measure of entanglement is the sum of the negative
eigenvalues (which will be zero if the system is not entangled). In a fermionic many-body system there
is always entanglement. However, as we have seen the extent of the entanglement changes as
we turn on the superconductivity, introduce a magnetic impurity and change the spin coupling to the
electronic spin density. We may expect therefore also some signature of the various transitions
between these physical states and systems in the negativity. In our case we actually have a $N$ site system
(and a number of electrons given by the band-filling determined by the chemical potential).
We construct a reduced two-site matrix integrating over $N-2$ site degrees of freedom.
These two sites, called above $i$ and $j$ may now take the role of A and B in the definition
of the negativity. We can therefore define a corresponding transposed reduced density matrix
where the basis states of for instance the $i$ site (A subsystem) are transposed. We can relate
the matrix elements of this transposed matrix to the matrix elements of the original
reduced density matrix. We can as well determine its eigenvalues and calculate the negativity.

\begin{figure}
\includegraphics[width=0.7\textwidth]{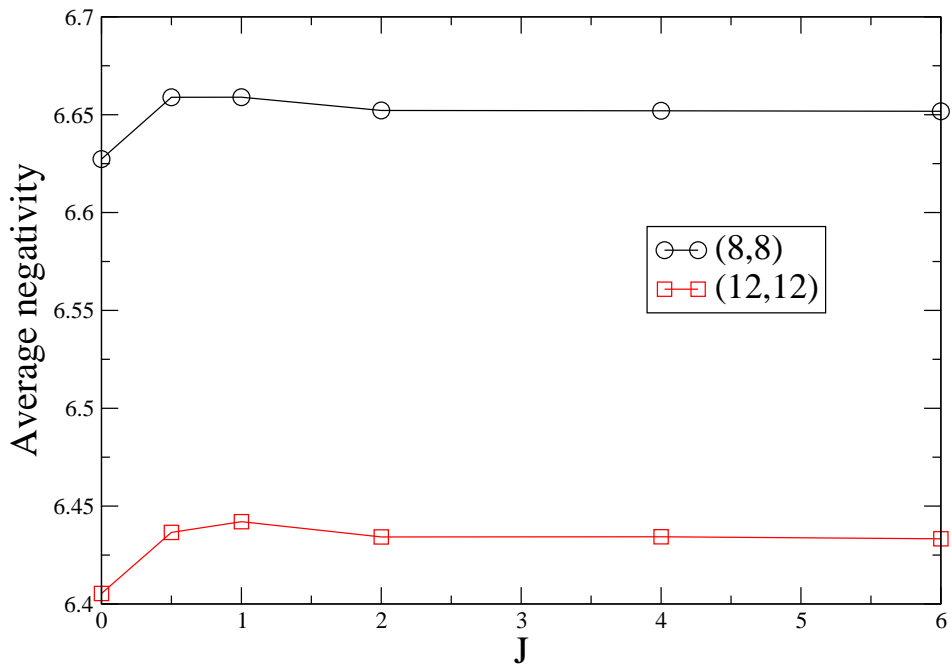}
\caption{\label{negat}
Average negativity as a function of $J$ where in one case one of the points is fixed at the impurity location and the other at
a site in the bulk.
}
\end{figure}

The negativity turns out to oscillate throughout the system. As we saw above, we could use the two-site
entanglement entropy to detect the phase transition due to the effect of the transition on the
entropy near the impurity location. However, as we saw this feature is small. While it is easily
detectable in the case of the two-site entanglement entropy, the oscillations in the negativity
are of the same order of magnitude and therefore not visible. However, as we saw the transition
also affects the background value. We may therefore consider the average negativity throughout the
system. This is shown in Fig. \ref{negat} where two cases are considered: one with the impurity as the fixed site and another with the fixed site in the bulk,
both as a function of the spin coupling.
In both cases we see that there is a slight change when the impurity is introduced
($J \neq 0$ with respect to the case of $J=0$) and when the critical point is crossed. However
the signature is quite small.

\section{Meyer-Wallach measure or generalized global entanglement}

Another measure of entanglement is the so-called generalized global entanglement cited recently
\cite{miranda} as particularly suited to detect a QPT. It is defined like
\be
G(2,n) = \frac{d}{d-1} \left[ 1 - Tr \left( \rho^2_{i,i+n} \right) \right]
\ee
where $d$ is the dimensionality of the reduced density matrix (in our case $d=16$) and the
two-site density matrix refers to two points $i$ and $j=i+n$. This quantity can be understood as the
deviation of the reduced density matrix from the one of a pure state, where it would be zero.
As referred in ref. \cite{miranda} this measure is expected to be at least as good
as the two-site entanglement entropy, analyzed above. The results for the function $G(2,n)$ are
qualitatively the same as those for the two-site entropy, and we will not present them here
in detail. When one of the sites, say $i$, is located at the impurity or at a bulk site the
same features are found. When one of the sites is at the impurity site and the QPT is crossed there
is a decrease in the background value and when one of the sites is located in the bulk
for a value smaller than the critical value of $J$ there is a small increase of the $G(2,n)$
at the impurity location and after the QPT is crossed there is a depletion, as for the two-site entropy.

\begin{figure}
\includegraphics[width=0.6\textwidth]{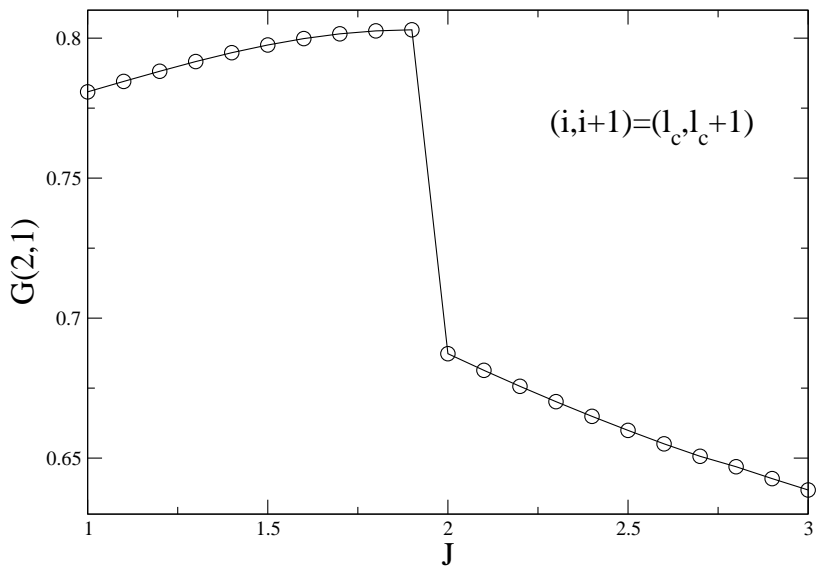}
\caption{\label{last}
Meyer-Wallach measure $G(2,1)$ as a function of $J$ where one of the sites 
is at the impurity location and the other at a nearest
neighbor. The QPT is clearly signaled.
}
\end{figure}

It is however a good way to detect the phase transition if the results are presented in a clearer way
(a similar role is achieved by the two-site entropy). Consider first the function $G(2,1)$ and take
$i=l_c=(8,8)$. In Fig. \ref{last} we show the effect of changing the spin coupling $J$ on this function.
The QPT is clearly identified.

\begin{figure}
\includegraphics[width=0.8\textwidth]{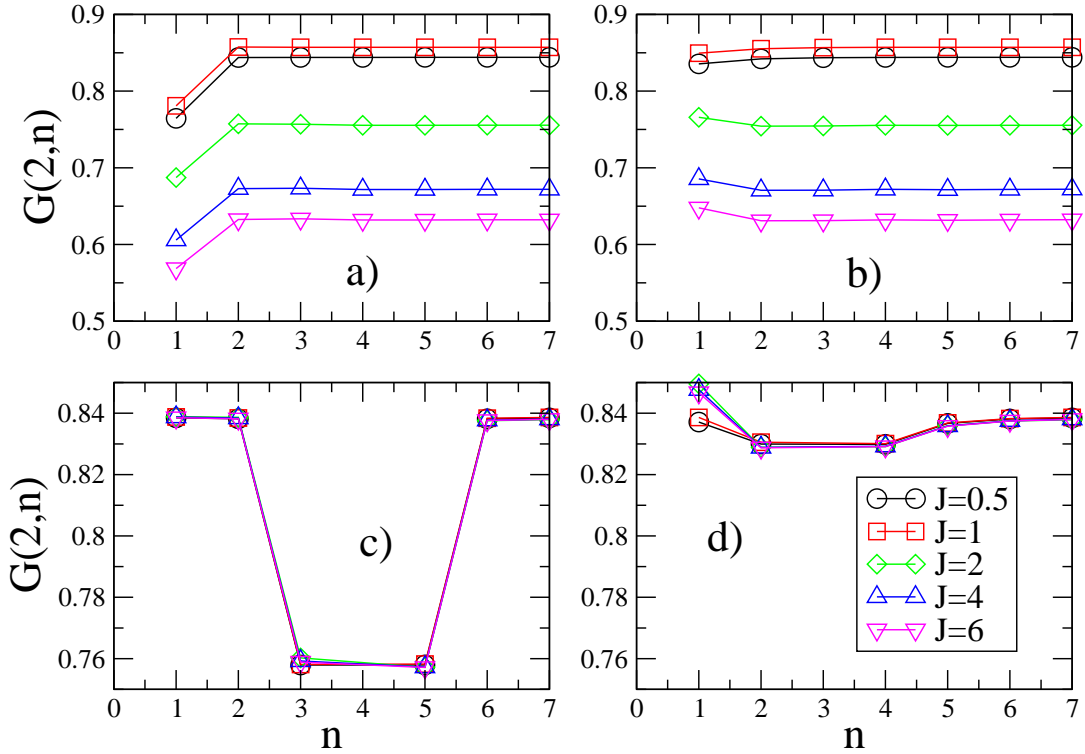}
\caption{\label{mirn}
$G(2,n)$ as a function of $n$ for different spin couplings: the point $i$ is fixed at the impurity location
and a) $n$ runs along the positive $x$ direction and b) along the upper side of the diagonal $z=x$, and
the point $i$ is fixed at a bulk site located along the same diagonal with c) $n$ along a horizontal line
to the right of the impurity and d) the same diagonal from the impurity location to the right upper corner of the
lattice.
}
\end{figure}

In Fig. \ref{mirn} we show the dependence with distance of the generalized global entanglement function $G(2,n)$,
where $n$ is chosen along various directions fixing one of the points at the impurity location or a site in
the bulk, for different spin couplings. 

\begin{figure}
\includegraphics[width=0.7\textwidth]{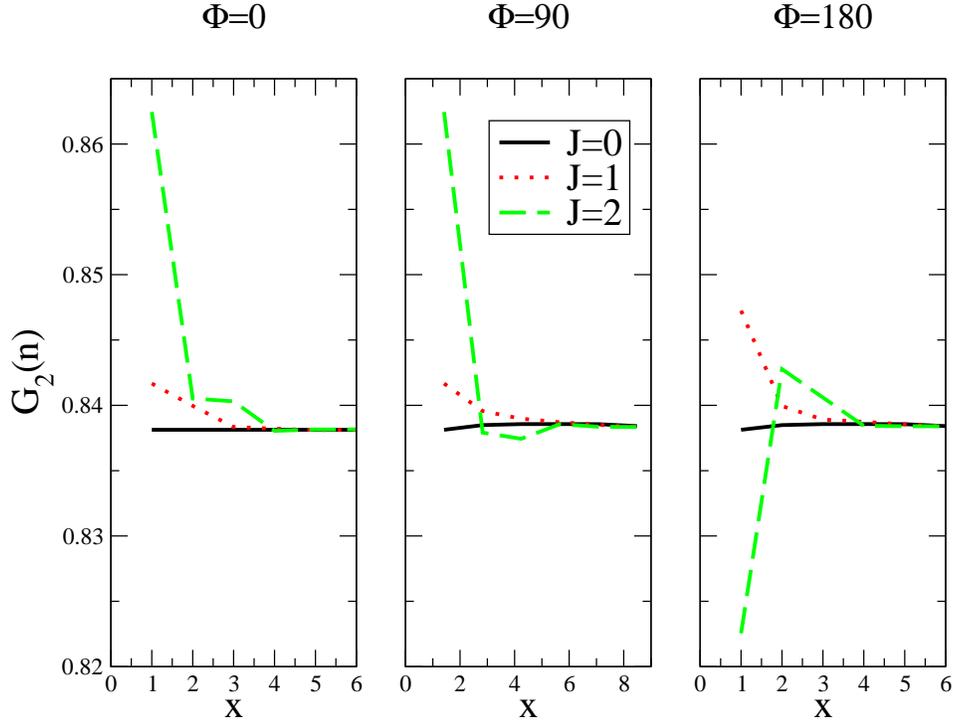}
\caption{\label{angles}
$G(2,n)$ as a function of the distance to the impurity site for different spin couplings. 
The angle $\Phi$ is defined between the two lines over which the two sites involved in
the entanglement are changing. The angle $\Phi=0$ defines two parallel lines along the horizontal
axis such that the distance between the two sites is fixed (in this case two lattice constants) but the
distance of the pair to the impurity, $x$ is increasing. The case $\Phi=\pi/2$ describes a situation where
one of the sites is moving along the diagonal $z=x$ and the other along the diagonal $z=-x$ and finally the
case $\Phi=\pi$ describes the case when one of the sites is moving along the vertical axis upward and
the other point along the same axis but downward. Therefore, in the cases $\Phi=\pi/2,\pi$ the distance
between the two points increasing while in the case $\Phi=0$ it is constant. In all cases the distance $x$ is
the distance between one of the members of the pair and the impurity site. 
}
\end{figure}

We conclude presenting the results for the same function $G(2,n)$ where the two sites $i$ and $i+n$ define two lines
that make an angle $\Phi=0,\pi/2, \pi$ between themselves. We compare the cases where there is no impurity ($J=0$)
with the insertion of the impurity $J=1,2$. The influence of the impurity is noticed for small distances between
the two sites and is particularly noted for the larger coupling. Interestingly, when the angle is large $\Phi=\pi$,
which corresponds to the two sites involved in the entanglement being along the z axis with the impurity
between them, the crossing of the QPT leads to a significant decrease of the function $G(2,n)$. This is shown
in Fig. \ref{angles}.

\section{Summary}

In this work we have shown that various entanglement measures are suited to detect
the quantum phase transition induced by a magnetic impurity inserted in a conventional
s-wave superconductor. The magnetic impurity induces a pair of bound states that go through
a level crossing as the coupling of the electronic spin density to the impurity spin grows.
In this quantum phase transition the impurity locks an electron, thereby locally changing the
spin density and the gap function, but it also changes the overall magnetization in a discontinuous
manner. This transition is also easily detected for instance in the von Neumann entropy
both for single-site and two-site subsystems. We have also considered the mutual information,
the negativity, and the recently introduced generalized global entanglement. All quantities
are sensitive to the change of the ground state of the system except for the negativity
which is less accurate.

Since in the problem studied in this paper the single-site von Neumann entropy is separable into a spin part
and a charge part (coupling empty sites and doubly occupied sites) we were able to show
that the spin part is much more affected at the QPT, as expected. In agreement with the
nature of the QPT, where an electron is captured by the impurity, the entropy at the impurity site
decreases and since the physical state approaches a pure state the generalized global entanglement
also decreases. Starting from a normal system (ie one with no impurities) and crossing the second order transition
to the superconducting phase naturally leads to a decrease in entropy. The insertion of the impurity has a
pair breaking effect and the entropy is locally increased. When the QPT is crossed the entropy
decreases locally in the vicinity of the impurity, as stated above. We have also shown
that the impurity and other sites are entangled, in the sense that the two-site von Neumann
entropy is slightly lower when the impurity site is involved, irrespective of the distance
between the impurity and the other site. However, this quantity is not a true measure of the
direct correlation between the two sites. It is actually a measure of the entanglement of the
two sites and the rest of the system. A better measure is the mutual information which
is sensitive to the correlations between the two sites. This quantity has a rather short
spatial range.

\begin{acknowledgments}
We acknowledge discussions with N. Paunkovic.
This work is supported by FCT Grant
No.~POCI/FIS/58746/2004 in Portugal, 
the ESF Science Programme INSTANS 2005-2010,
Polish Ministry of Science and Higher Education as a research
project in years 2006-2009,
and by STCU Grant No. 3098 in Ukraine. 
\end{acknowledgments}

\end{document}